\documentclass[journal]{IEEEtran}

\usepackage{graphicx}
\usepackage{wrapfig}
\usepackage{graphicx}
\usepackage{subfigure}
\usepackage{float}
\usepackage{adjustbox}

\usepackage[justification=centering]{caption}
\usepackage{tabularx}
\usepackage{multirow}
\usepackage[belowskip=-5pt,aboveskip=0pt]{caption}
\usepackage{array}
\newcolumntype{P}[1]{>{\centering\arraybackslash}p{#1}}
\newcolumntype{M}[1]{>{\centering\arraybackslash}m{#1}}
\newcommand{\PreserveBackslash}[1]{\let\temp=\\#1\let\\=\temp}
\newcolumntype{R}[1]{>{\PreserveBackslash\raggedleft}p{#1}}
\usepackage{tabulary,booktabs}
\usepackage{lipsum}
\usepackage{blindtext}
\usepackage{amsmath,amssymb}

\usepackage{hyperref}

\usepackage{enumitem}

\usepackage[linesnumbered, algo2e, ruled,norelsize]{algorithm2e}
\usepackage{xcolor}
\usepackage{cite}

\SetCommentSty{mycommfont}

\usepackage{gensymb}
\begin{document}
\title{Lossless Coding of Point Cloud Geometry using a Deep Generative Model}

\author{Dat Thanh Nguyen, Maurice Quach,~\IEEEmembership{Student Member,~IEEE,} Giuseppe Valenzise,~\IEEEmembership{Senior Member,~IEEE}, Pierre Duhamel,~\IEEEmembership{Life Fellow, IEEE}% <-this % stops a space
\thanks{D. T. Nguyen, M. Quach, G. Valenzise and P. Duhamel are with the Université Paris-Saclay, CNRS, CentraleSupelec, Laboratoire des Signaux et Systèmes (UMR 8506), 91190 Gif-sur-Yvette, France (email: \href{mailto:doandat.nguyen@gmail.com}{thanh-dat.nguyen@centralesupelec.fr}; \href{mailto:maurice.quach@l2s.centralesupelec.fr}{maurice.quach@l2s.centralesupelec.fr}; \href{mailto:giuseppe.valenzise@l2s.centralesupelec.fr}{giuseppe.valenzise@l2s.centralesupelec.fr}; \href{mailto:pierre.duhamel@l2s.centralesupelec.fr}{pierre.duhamel@l2s.centralesupelec.fr}).}}% <-this % stops a space


\maketitle

\begin{abstract}
This paper proposes a lossless point cloud (PC) geometry compression method that uses neural networks to estimate the probability distribution of voxel occupancy. First, to take into account the PC sparsity, our method adaptively partitions a point cloud into multiple voxel block sizes. This partitioning is signalled via an octree. Second, we employ a deep auto-regressive generative model to estimate the occupancy probability of each voxel given the previously encoded ones. We then employ the estimated probabilities to code efficiently a block using a context-based arithmetic coder. Our context has variable size and can expand beyond the current block to learn more accurate probabilities. We also consider using data augmentation techniques to increase the generalization capability of the learned probability models, in particular in the presence of noise and lower-density point clouds. Experimental evaluation, performed on a variety of point clouds from four different datasets and with diverse characteristics, demonstrates that our method reduces significantly (by up to 30\%) the rate for lossless coding compared to the state-of-the-art MPEG codec.
%, with a significant improvement on point clouds taken from Microsoft Voxelized Upper Bodies (MVUB), 8i, MPEG, and University of Sao Paulo (USP) datasets,  respectively. 
%Within each voxel block, geometric information is preserved and can be processed by a neural network.
\end{abstract}


\section{Introduction}

\label{sec:intro}

\IEEEPARstart{P}{oint} clouds (PC) are becoming the most popular data structure for many 3D applications such as augmented, mixed or virtual reality, as they enable six degrees of freedom (6DoF) interaction. 
%including three translation and three rotation movements.
Typical PCs contain millions of points, each point being represented by $x,y,z$ coordinates, and attributes (e.g. color, normal, etc.). This entails a high transmission and storage cost. As a result, there is a massive demand for efficient Point Cloud Compression (PCC) methods to enable the practical use of this content. 
\par %The geometry and attributes of PCs are usually independently encoded. 
The Moving Picture Expert Group (MPEG) has studied coding solution for various categories of point clouds, including static point clouds (category 1), dynamic point clouds (category 2), and LiDAR sequences (category 3 -- dynamically acquired point clouds). As a result, two PCC standards have been developed \cite{8571288,jang2019video,graziosi2020overview}: Video-based PCC (V-PCC) and Geometry-based PCC (G-PCC). V-PCC focuses on dynamic point clouds, and projects the volumetric video onto 2D planes before encoding. The generated 2D videos are then compressed using 2D video coding standards. This approach benefits from efficient 2D video coding solutions which have been optimized over several decades.  On the other hand, G-PCC targets static content, and the geometry and attribute information are independently encoded. Color attributes can be encoded using methods based on the Region Adaptive Hierarchical Transform (RAHT) \cite{de2016compression}, Predicting Transform or Lifting Transform \cite{graziosi2020overview}. 
Coding the PC geometry is particularly important to convey the 3D structure of the PC, but is also challenging, as the non-regular sampling of point clouds makes it difficult to use conventional signal processing and compression tools.
% However, the geometry must be available before filling point clouds with attributes. 
In this paper, we focus on \textit{lossless coding} of point cloud geometry. 
\par In particular, we consider the case of \textit{voxelized} point clouds. Voxelization consists in pre-quantizing the geometric coordinates of the point cloud prior to coding in order to represent the geometry with integer precision. This operation is common in many coding scenarios, e.g., when dealing with dense point clouds such as those produced by camera arrays.
After voxelization, the point cloud geometry can be represented either directly in the voxel domain or using an octree spatial decomposition. PCs are divided into a fixed number of cubes, which defines the resolution (e.g., 10 bit = 1024 cubes per dimension).  Each cube is called a voxel. If a voxel contains at least one point, it is called an occupied voxel. Usually, very few voxels are occupied and a large part of the volume is empty. 
% In this paper, we also adopt this approach, which is particularly suited for dense point clouds. 
An octree representation can be obtained by %analyzing the geometry of the voxelized point cloud. Assuming that the point cloud is mapped into a volume of $D \times D \times D$ voxels, the volume is 
recursively splitting the volume into eight sub-cubes until the desired precision is achieved. Then, occupied blocks are marked by bit 1 and empty blocks are marked by bit 0. Consequently, at each level, the generated 8 bits represent the occupancy state of an octree node (octant). Our method operates in both the voxel and octree domain. On the one hand, the octree representation can naturally adapt to the  sparsity of the point cloud, as empty octants do not need to be further split; on the other hand, in the voxel domain convolutions can be naturally expressed, and geometric information (i.e., planes, surfaces, etc.) can be explicitly processed by a neural network. 

% \par MPEG identified three categories of point clouds: static point clouds (category 1), dynamic point clouds (category 2), and LiDAR sequences (category 3 - dynamically acquired point clouds). 
In this work, we propose a deep-learning-based method (named VoxelDNN) for lossless compression of static voxelized point cloud geometry. Our main contributions are:
\begin{itemize}
    \item We employ for the first time a deep generative model in the voxel domain to estimate the occupancy probabilities sequentially using a masked 3D convolutional network. The conditional distribution is then used to model the context of a context-based arithmetic coder.
    \item We propose an optimal rate-driven partitioning and context selection algorithm. The partitioning algorithm adapts to the point cloud sparsity by employing a hybrid octree/voxel representation while the context to encode each block is expanded to the neighboring blocks and the expansion size is optimally selected.
    \item  We propose specific data augmentation techniques for 3D point clouds coding, to increase its generalization capability.
\end{itemize}
     
% The encoded PC bitstream consists of the partitioning signal in the form of octree, which is sent to the decoder as side information, as well as the occupied voxel blocks entropy-encoded using the distributions estimated by our context model.
%    \item We expand the context to the encoded voxels in the neighboring blocks to obtain a better probability estimation. The expansion size is adaptively selected during the block partitioning process.
\begin{figure*}
\centering
\captionsetup{justification=raggedright}
% \captionsetup{singlelinecheck = false, format= hang, justification=raggedright, font=small, labelsep=space}
\includegraphics[width=.95\linewidth]{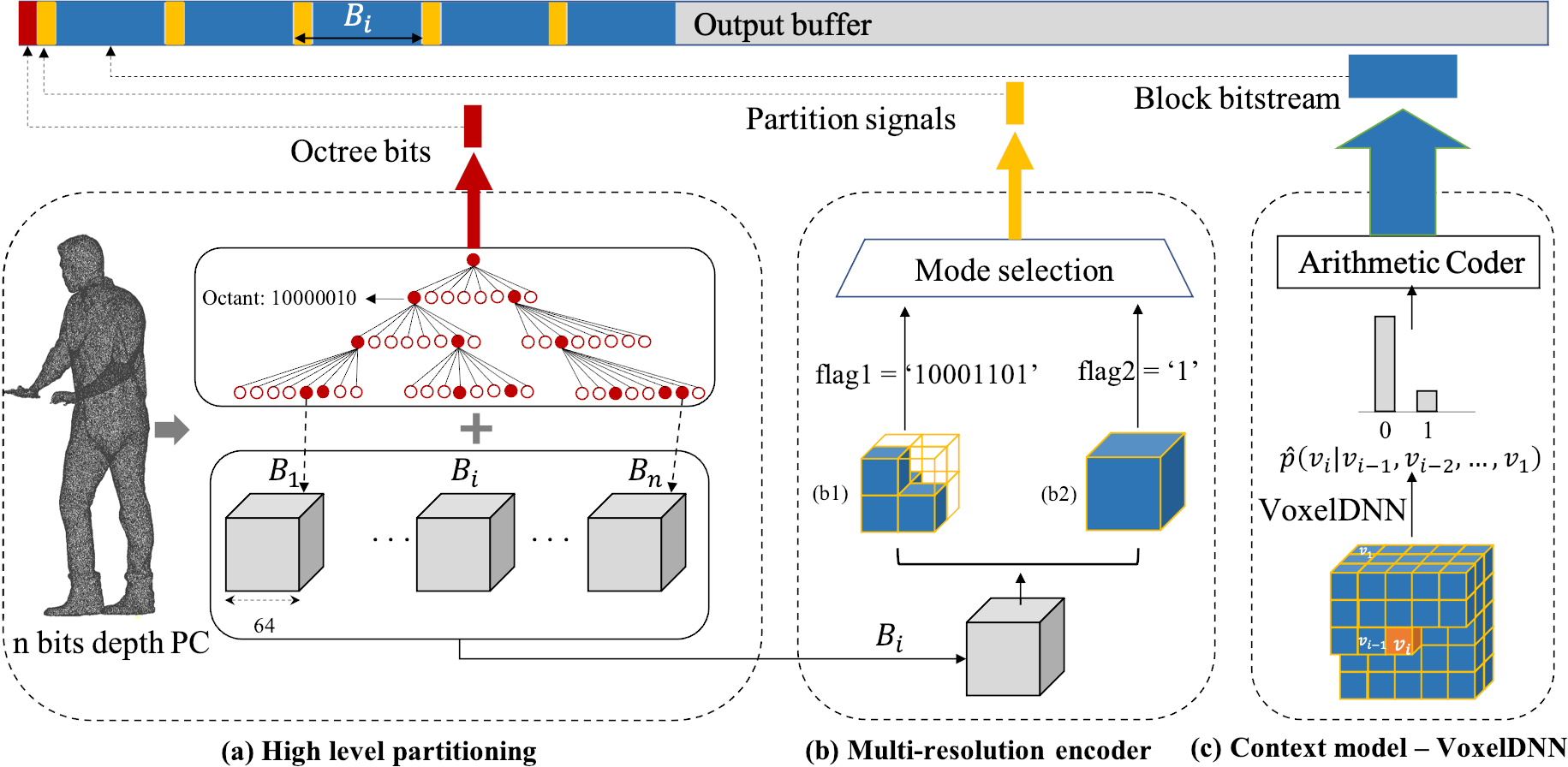}
\vspace{0.1cm}
\caption{Overview of the proposed method. (a): a $n$ bit depth point cloud is partitioned down to the $n-6$ octree level, yielding occupied blocks of size $64\times 64 \times 64$. (b): We encode each block of $64^3$ voxels as a single block (b1), or divide it into 8 children blocks (b2), depending on the total number of bits of each solution (partitioning level = 2). This procedure is repeated recursively for increasing partitioning levels up to 5. (c): For each occupied block of size $d $, the context model estimates the distribution of each voxel given the previously encoded voxels.  }
\label{fig:system overview}
\end{figure*}

\par We demonstrate experimentally that the proposed solution
outperforms the state-of-the-art MPEG G-PCC lossless codec in terms of bits per occupied voxel over a set of point clouds with varying density and content type. The rest of the paper is structured as follows: Section \ref{sec:stateoftheart} reviews the related work; the proposed method is described in Section \ref{proposedmethod}; Section \ref{performanceeval} presents the experimental results; and finally Section \ref{conclusion} concludes the paper.

\section{Related work}
\label{sec:stateoftheart}

\par Relevant work related to this paper includes state-of-the-art PC geometry coding and learning-based methods in image and point cloud compression.

\subsection{MPEG G-PCC and Conventional Lossless Codecs}
\par Most existing methods that compress point cloud geometry, including MPEG G-PCC, use octree coding \cite{schnabel2006octree,7434610,6224647, garcia2017context, garcia2018intra, garcia2019geometry,huang2020octsqueeze, biswas2020muscle} and local approximations called ``triangle soups'' (trisoup)~\cite{schnabel2006octree,dricot2019adaptive}. 
\par In the G-PCC geometry coder, points are first transformed and voxelized into an axis-aligned bounding box before geometry analysis using trisoup or octree scheme. In the trisoup coder, geometry can be represented by a pruned octree plus a surface model. This model approximates the surface in each leaf of the pruned octree using 1 to 10 triangles. 
% This technique is known as triangle soup. 
In contrast, the octree coder partitions voxelized blocks until sub-cubes of dimension one are reached. First, the coordinates of isolated points are independently encoded to avoid "polluting" the octree coding (Direct Coding Mode - DCM) \cite{dcm}. To encode the occupancy pattern of each octree node, G-PCC introduces many methods to exploit local geometry information and obtain an accurate context for arithmetic coding, such as Neighbour-Dependent Entropy Context \cite{neighbor}, intra prediction \cite{intracodinggpcc}, planar/angular coding mode \cite{planarcodingmode,angularcodingmode}, etc. The lossless geometry coding mode of G-PCC is based on octree coding only.

\par In order to deal with the irregular point space, many octree-based lossless PCC methods have been proposed. In \cite{schnabel2006octree}, the authors proposed an octree-based method which aims at reducing entropy by employing prediction techniques based on local surface approximations to predict occupancy patterns. Recently, more context modeling based approaches are proposed \cite{garcia2017context, garcia2018intra, garcia2019geometry}. For example, the intra-frame compression method P(PNI) proposed in \cite{garcia2019geometry} builds a reference octree by propagating the parent octet to all children nodes, thus providing 255 contexts to encode the current octant. Octree coding allows for a progressive representation of point clouds since each level of the octree is a downsampled version of the point cloud. However, a drawback of octree representation is that, at the first levels of the tree, it produces ``blocky'' scenes, and geometry information of point clouds (i.e., curve, plane) is lost. The authors of \cite{8122226} proposed a binary tree based method which analyzes the point cloud geometry using binary tree structure and realizes an intra prediction via the extended Travelling Salesman Problem (TSP) within each leaf node. Instead, in this paper, we employ a hybrid octree/voxel representation to better exploit the geometry information. Besides,  the methods in \cite{garcia2017context, garcia2018intra, garcia2019geometry} produce frequency tables which are collected from the coarser level or the previous frame and must be transmitted to the decoder. Our method predicts voxel distributions in a sequential manner at the decoder side, thus avoiding the extra cost of transmitting large frequency tables.
%  The lossless geometry compression method of \cite{6224647} is based on predictive coding with inter-frame prediction
\subsection{Generative Models and Learning-based Compression}
\par 
% Learning probabilistic models that return explicit probability densities from training data is the central problem in unsupervised learning.
Estimating the data distribution from a training dataset is the main objective of generative models, and is a central problem in unsupervised learning.
It has a number of applications, from image generation \cite{theis2015generative,gregor2015draw, oord2016pixel,salimans2017pixelcnn++}, to image compression \cite{491334, balle2016end,mentzer2018conditional} and denoising \cite{chen2018image}. Among the several types of generative models proposed in the literature \cite{10.5555/2969033.2969125}, auto-regressive models such as PixelCNN \cite{oord2016pixel,salimans2017pixelcnn++} are particularly relevant for our purpose as they allow to compute the exact likelihood of the data and to generate realistic images, although with a high computational cost. Specifically, PixelCNN factorizes the likelihood of a picture by modeling the conditional distribution of a given pixel's color given all previously generated pixels. These conditional distributions only depend on the possible pixel values with respect to the scanned context, which imposes a \textit{causality} constraint. PixelCNN models the distribution using a neural network and the causality constraint is enforced using masked filters in each convolutional layer. Recently, this approach has also been employed in image compression to yield accurate and learnable entropy models~\cite{mentzer2018conditional}. Our paper explores the potential of this approach for point cloud geometry compression by adopting and extending conditional image modeling and masking filters into the 3D voxel domain.

\par Inspired by the success in learning-based image compression, deep learning has been widely adopted in point cloud coding  both in the octree domain \cite{huang2020octsqueeze,biswas2020muscle},  voxel domain \cite{8954537,9191021,quach2019learning,quach2020improved,wang2019learned,guarda2020point}  and point domain \cite{yan2019deep,huang20193d, wang2020multiscale}. Recently, the authors of \cite{huang2020octsqueeze} proposed an octree-based entropy model that models the probability distributions of the octree symbols based on the contextual information from octree structure. This method only targets static LiDAR point cloud compression. The extension version for intensity-valued LiDAR streaming data using spatio-temporal relations is proposed in \cite{biswas2020muscle}. However, these methods target dynamically acquired point clouds, while in this paper we mainly focus on dense static point clouds.
\par Working in the voxel domain enables to easily extend most 2D tools, such as convolutions, to the 3D space. 
Many recent 3D convolution based autoencoder approaches for lossy coding \cite{quach2019learning,quach2020improved,wang2019learned,guarda2020point} compress 3D voxelized blocks into latent representations and cast the reconstruction as a binary classification problem. The authors of \cite{yan2019deep} proposed a pointnet-based auto-encoder method which directly takes points as input rather than voxelized point cloud.  To handle sparse point clouds, recent methods leverage advances in sparse convolution \cite{choy20194d,graham2017submanifold} to allow point-based approaches  \cite{huang20193d, wang2020multiscale}. For example, the proposed lossy compression method in \cite{wang2020multiscale} progressively downscale the point cloud into multiple scales using sparse convolutional transforms. Then, at the bottleneck, the geometry of scaled point cloud is encoded using an octree codec and the attributes are compressed using a learning-based context model.  In contrast, in this paper, we focus on dense voxelized point clouds and losslessly encode each voxel using  the learned distribution from its 3D context. In addition, we apply this approach in a block-based fashion, which has been successfully employed in traditional image and video coding.
%and tackle the sparsity by representing point cloud in hybrid octree/voxel domain

\section{Proposed method}
\label{proposedmethod}
\subsection{System overview}
\label{problem fomulation}
In this work, we propose a learning-based method for lossless compression of point cloud geometry. We aim at minimizing the encoded rate measured by the number of bits per occupied voxel (bpov) by exploiting the spatial redundancies within point cloud. The general scheme of our method is shown in Figure \ref{fig:system overview}. A point cloud voxelized over a $2^n \times 2^n \times 2^n$ grid is known as an $n$-bit depth PC, which can be represented by an $n$ level octree. In this work, we represent point cloud geometry in a hybrid manner, by combining  the octree and voxel domains. We coarsely partition an $n$-depth point cloud up to  level $n-6$. This allows to coarsely remove most of the empty space in the point cloud. As a result, we obtain a $n-6$ level octree and a number of non-empty binary blocks $v$ of size $2^6 \times 2^6 \times 2^6$ voxels, which we refer to as resolution $d=64$ or simply block 64 (Figure \ref{fig:system overview}(a)). Blocks 64 can be further partitioned at resolution $d=\{64,32, 16, 8, 4\}$ corresponding to maximum partitioning level $maxLv=\{1,2,3,4,5\}$ as detailed in Section~\ref{ssec:multires}. Figure \ref{fig:system overview}(b) shows the multi-resolution encoder with $maxLv=2$.  A block of size $d$ can be encoded as a single block (b2) or partitioned into 8 sub-cubes (b1). We then encode each non-empty block (blocks in blue in the figure) using the proposed method in the voxel domain (Section~\ref{ssec:voxelDNN}) and select the  partitioning mode resulting in the smallest bpov. The overview of a single block encoder is shown in Figure \ref{fig:system overview}(c). Our context model predicts the distribution of each voxel given all encoded voxels and pass it to an arithmetic coder to generate the final bitstream. The context is chosen adaptively following a rate optimization algorithm (Section~\ref{ssec:multires}). The high-level octree, partitioning signal, selected context as well as the depth of each block are converted to bytes and signaled to the decoder as side information. For ease of notation, we index all voxels in block $v$ at resolution $d$ from $1$ to ${d^3}$ in raster scan order with:
 \begin{equation}
    v_i= 
    \begin{cases} 
    1, \quad \text{if $i^{th}$ voxel is occupied}\\
    0, \quad \text{otherwise}.
    \end{cases}
\label{focalloss}
\end{equation}

%\par Point cloud data structure, voxel domain, octree domain
\subsection{VoxelDNN}\label{ssec:voxelDNN}

\begin{figure}
\captionsetup{justification=raggedright}
%\captionsetup{singlelinecheck = false, format= hang, justification=raggedright, font=small, labelsep=space}
\begin{minipage}[b]{.45\linewidth}
  \centering
  \centerline{\includegraphics[width=0.85\linewidth]{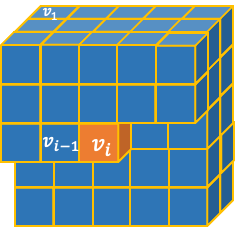}}
%  \vspace{1.5cm}
  \centerline{(a) 3D voxel context}\medskip
\end{minipage}
\hfill
\begin{minipage}[b]{0.45\linewidth}
\label{sfig:typeA}
  \centering
  \centerline{\includegraphics[width=0.95\linewidth]{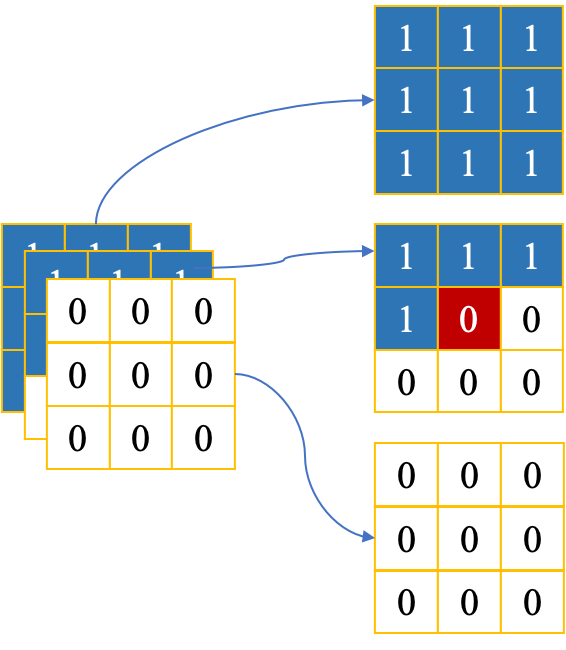}}
%  \vspace{1.5cm}
  \centerline{(b) 3D type A mask }\medskip
\end{minipage}
\caption{(a): Example 3D context in a $5 \times 5 \times 5$ block. Previously scanned elements are in blue. (b): $3 \times 3 \times 3$ 3D type A mask. Type B mask is obtained by changing center position (marked red) to 1. }
\label{fig:context}
\end{figure}
\begin{figure}[b]
\captionsetup{justification=raggedright}
\includegraphics[width=0.9\linewidth]{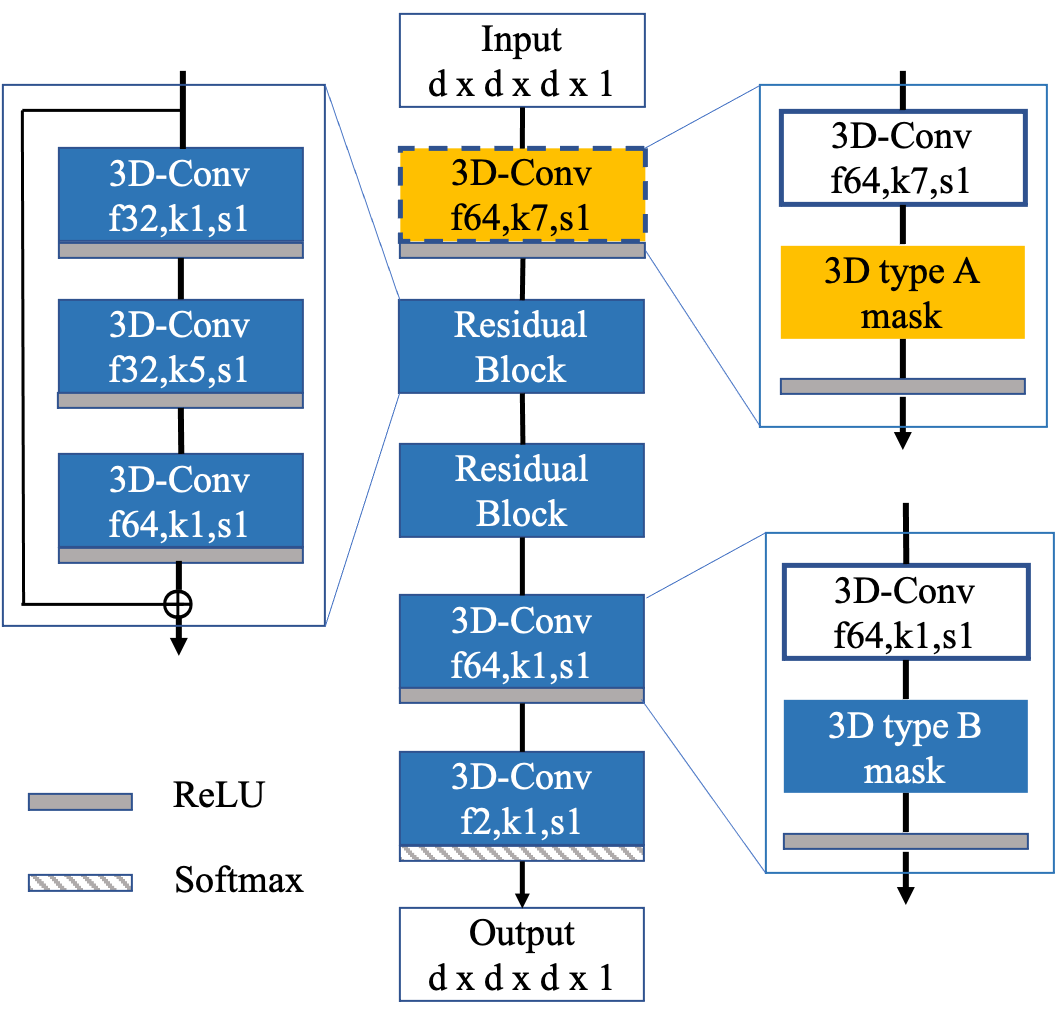}
\caption{VoxelDNN architecture, $d$ is the dimension of the input block, masked layers are colored in yellow and blue. A type A mask is applied to the first layer (dashed borders) and type B masks afterwards. `f64,k7,s1' stands for 64 filters, kernel size 7 and stride 1. }
\label{fig:Networkarchitecture}
\end{figure}

\begin{algorithm2e}[t]
\footnotesize
\SetAlgoLined
\SetKwInOut{Input}{Input}
\SetKwInOut{Output}{Output}
\KwIn{block: $B$, current level: $curLv$, max level: $maxLv$} 
\KwOut{partitioning flags: $fl$, output bitstream: $bits$}
\SetKwFunction{FMain}{partitioner}
\SetKwProg{Fn}{Function}{:}{}
\Fn{\FMain{$B, curLv, maxLv$}}{
    %\tcc{encode as 8 child blocks}
    $fl2$ $\leftarrow$ 2  \tcp*[l]{encode as 8 child blocks}
    \For{block $b$ in child blocks of $B$}{
        \eIf{$b$ is empty}{
            $child\_flag$ $\leftarrow$ 0\;
            $child\_bit$ $\leftarrow empty$\;
        }{
            \eIf{$curLv$ $==$ $maxLv$}{
                %\tcc{encode child block as a single block}
                $child\_flag$ $\leftarrow$ 1\;
                $child\_bit$ $\leftarrow$ singleBlockCoder($b$)\;
            }{
                %\tcc{paritioning selection at lower level}
                $child\_flag, child\_bit$ $\leftarrow$ partitioner($b$, $curLv+1$,$maxLv$)\;
            }
        }
        $ fl2 \leftarrow$ [$fl2$, $child\_flag$]\;
        $ bit2\leftarrow$ [$bit2$, $child\_bit$]\; 
    }
    $total\_bit2=sizeOf(bit2)+len(fl2)\times 2$\;
    %\tcc{encode as a single block}
    $fl1 \leftarrow$ 1\tcp*[l]{encode as a single block}
    $bit1 \leftarrow$ singleBlockCoder($B$)\;
    $total\_bit1=sizeOf(bit1)+len(fl1)\times 2$\;
    \tcc{partitioning selection}
    \eIf{$total\_bit2 \geq total\_bit1$}{
        \KwRet $fl1, bit1$\;
    }{
        \KwRet $fl2, bit2$\;
    }
  }
\caption{Block partitioning selection}
\label{algo:proposed_method}
\end{algorithm2e}
%\setlength{\textfloatsep}{0pt}

% \subsubsection{Voxel context model}
\par Our method losslessly encodes the voxelized point cloud  using context-adaptive binary arithmetic coding. Specifically, we focus on estimating accurately a probability model $p(v)$  for the occupancy of a block $v$ composed by $d \times d \times d$ voxels. We factorize the joint distribution $p(v)$ as a product of conditional distributions $p(v_i|v_{i-1}, \ldots, v_1)$ over the voxel volume: 
\begin{equation}
    p(v)= \underset{i=1 }{\overset{d^3}{\Pi}}p(v_i|v_{i-1},v_{i-2},\ldots,v_{1}).
    \label{eq:p(v)}
\end{equation}
Each term $p(v_i|v_{i-1}, \ldots, v_1)$ above is the probability of the voxel $v_{i}$ being occupied given the occupancy of all previous voxels,  referred to as a context. Figure \ref{fig:context}(a) illustrates such a 3D context. We estimate $p(v_i|v_{i-1}, \ldots, v_1)$ using a neural network which we dub \textbf{VoxelDNN}.  

\par The conditional distributions in~\eqref{eq:p(v)} depend on previously decoded voxels. This requires a \textit{causality} constraint on the VoxelDNN network. To enforce causality, we extend to 3D the idea of masked convolutional filters, initially proposed in PixelCNN~\cite{oord2016pixel}. Specifically, two kinds of masks (A or B) are employed. Type A mask is filled by zeros from the center position to the last position in raster scan order as shown in Figure \ref{fig:context}(b). Type B mask differs from type A in that the value in the center location is 1 (colored in red). Type A masks are used in the first convolutional filter to remove the connections between all future voxels and the voxel currently being predicted. From the second layer, the value of the current voxel is not used in its spatial position and is replaced by the result of the convolution over previous voxels.  As a result, from the second convolutional layer, type B masks are applied which relaxes the restrictions of mask A by allowing the connection from the current spatial location to itself. 

\par In order to learn good estimates $\hat{p}(v_i|v_{i-1}, \ldots, v_1)$ of the underlying voxel occupancy distribution $p(v_i|v_{i-1}, \ldots, v_1)$, and thus minimize the coding bitrate, we train VoxelDNN using cross-entropy loss. That is, for a block $v$ of resolution $d$, we minimize :
\begin{equation}\label{eq:CEloss}
    H(p,\hat{p}) = \mathbb{E}_{v\sim p(v)}\left[\sum_{i=1}^{d^3} -\log \hat{p}(v_i)\right].
\end{equation}
It is well known that cross entropy represents the extra bitrate cost to be paid when the approximate distribution $\hat{p}$ is used instead of the true $p$. More precisely, $H(p,\hat{p}) = H(p) + D_{KL}(p\| \hat{p})$, where $D_{KL}$ denotes the Kullback-Leibler divergence and $H(p)$ is Shannon entropy. Hence, by minimizing~\eqref{eq:CEloss}, we indirectly minimize the distance between the estimated conditional distributions and the real data distribution, yielding accurate contexts for arithmetic coding. Note that this is different from what is typically done in learning-based \textit{lossy} PC geometry compression, where the focal loss is used~\cite{quach2019learning, quach2020improved}. In this lossy context, the motivation behind using focal loss is to cope with the high spatial unbalance between occupied and non-occupied voxels. The reconstructed PC is then obtained by hard thresholding $\hat{p}(v)$, and the target is thus the final classification accuracy. Conversely, here we aim at estimating accurate soft probabilities to be fed into an arithmetic coder.

\par Figure \ref{fig:Networkarchitecture} shows our VoxelDNN network architecture for a block  of dimension $d$. Given the $d \times d \times d$ input block, VoxelDNN outputs the predicted occupancy probabilities of all input voxels. Our first 3D convolutional layer uses $7 \times 7 \times 7$ kernels with a type A mask. Type B masks are used in the subsequent layers. To avoid vanishing gradients and speed up the convergence, we implement two residual blocks \cite{he2016deep} with $5 \times 5 \times 5$ kernels. Since type A masks are applied at the first layer, identity skip connection of residual block does not violate the causality constraint. Throughout VoxelDNN, the ReLu activation function is applied after each convolutional layer, except in the last layer where we use softmax activation. Using more filters generally increases the performance of 
 VoxelDNN, at the expense of an increase in the number of parameters and computational complexity. After experimenting with various number of filters, we concluded that for input voxel block ($d \times d \times d \times 1$) which only has a single feature, 64 convolutional filters give a good trade-off between complexity and model performance.
\subsection{Multi-resolution encoder and adaptive partitioning}\label{ssec:multires}

We use an arithmetic coder to encode the voxels sequentially from the first voxel to the last voxel of each block in a generative manner. Specifically, every time a voxel is encoded, it is fed back into VoxelDNN to predict the probability of the next voxel. Then, we pass the probability to the arithmetic coder to encode the next symbol. 

\begin{figure}
\captionsetup{justification=raggedright}
\begin{minipage}[b]{.24\linewidth}
  \centering
  \centerline{\includegraphics[width=0.90\linewidth]{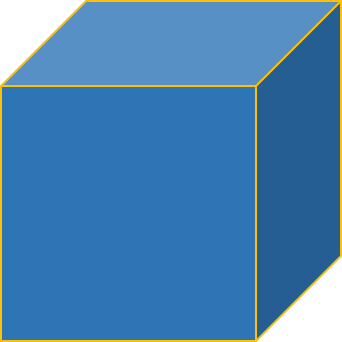}}
%  \vspace{1.5cm}
  \centerline{(a)}\medskip
\end{minipage}
\hfill
\begin{minipage}[b]{0.24\linewidth}
  \centering
  \centerline{\includegraphics[width=0.90\linewidth]{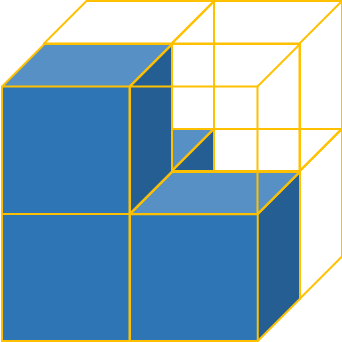}}
%  \vspace{1.5cm}
  \centerline{(b) }\medskip
\end{minipage}
\hfill
\begin{minipage}[b]{0.24\linewidth}
  \centering
  \centerline{\includegraphics[width=0.90\linewidth]{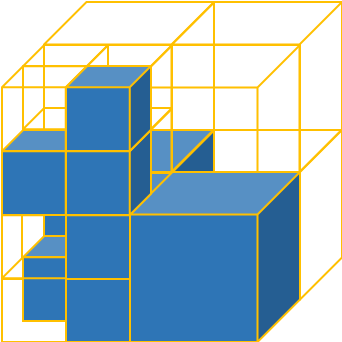}}
%  \vspace{1.5cm}
  \centerline{(c) }\medskip
\end{minipage}
\hfill
\begin{minipage}[b]{0.24\linewidth}
  \centering
  \centerline{\includegraphics[width=0.90\linewidth]{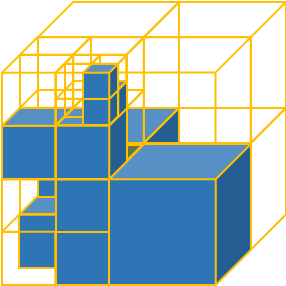}}
%  \vspace{1.5cm}
  \centerline{(d)}\medskip
\end{minipage}
\caption{Partitioning a block of size 64 into: (a) a single block of size 64, (b): blocks of size 32, (c): 32 and 16, (d): 32, 16 and 8. Non-empty blocks are indicated by blue cubes.}
\label{fig:4levelpartiitoning}
\end{figure}

However, applying this coding process at a fixed resolution $d$ (in particular, on larger blocks) can be inefficient when blocks are sparse, i.e.,  they contain only a few occupied voxels. 
% encoding the whole block as a single block is not always the best solution, especially when blocks are sparse. 
This is due to the fact that in this case, there is little or no information available in the receptive fields of the convolutional filters. To overcome this problem, we propose to optimize the block size based on a rate-optimized multi-resolution splitting algorithm as follows.
We partition a block into 8 sub-blocks recursively and signal the occupancy of sub-blocks as well as the partitioning decision (0: empty, 1: encode as a single block, 2: further partition). The partitioning decision depends on the bit rate after arithmetic coding. If the total bitstream of partitioning flags and occupied sub-blocks is larger than encoding the parent block as a single block, we do not perform partitioning. The details of this process are shown in Algorithm \ref{algo:proposed_method}. The maximum partitioning level or the maximum number of block sizes is controlled by $maxLv$ and partitioning is performed up to $maxLv=5$ corresponding to a smallest block size of 4. Depending on the output bits of each partitioning solution, a block of size 64 can contain a combination of blocks with different sizes. Figure \ref{fig:4levelpartiitoning} shows 4 partitioning examples for an encoder with $maxLv=4$. Note that VoxelDNN learns to predict the distribution of the current voxel based on previously encoded voxels. As a result, we can use a bigger model size to predict the probabilities for smaller input block size.

\subsection{Context extension}\label{ssec:extendcontext}
\begin{figure}[tb]
\captionsetup{justification=raggedright}
\begin{minipage}[b]{.45\linewidth}
  \centering
  \centerline{\includegraphics[width=0.90\linewidth]{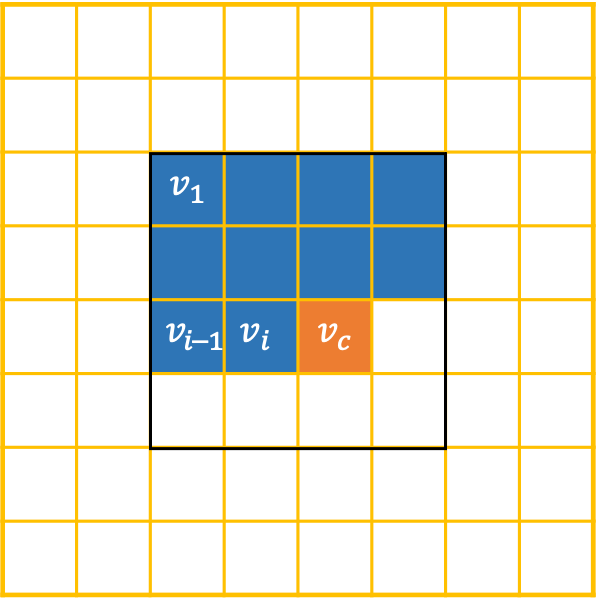}}
%  \vspace{1.5cm}
  \centerline{(a)}\medskip
\end{minipage}
\hfill
\begin{minipage}[b]{0.45\linewidth}
  \centering
  \centerline{\includegraphics[width=0.90\linewidth]{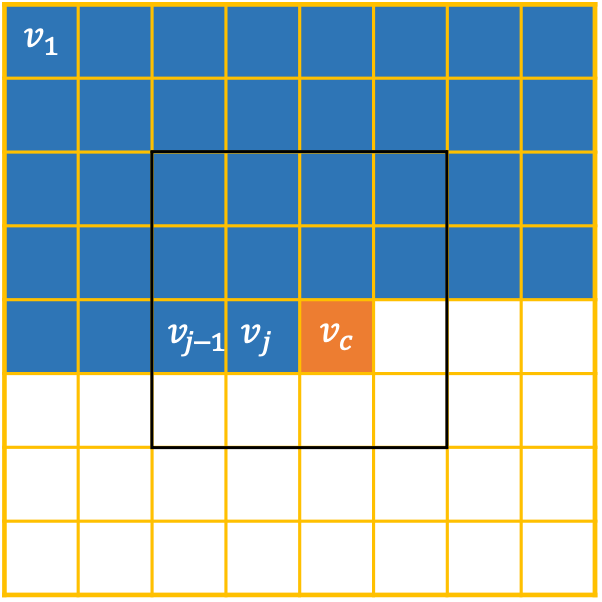}}
%  \vspace{1.5cm}
  \centerline{(b) }\medskip
\end{minipage}
\caption{2D illustration of context extension from block $4 \times 4$ to block $8 \times 8$. (a): Before extension, (b): after extension. Blue squares are active voxels in the context, voxels in the white area are ignored by masks or from the bigger block.  }
\label{fig:context extension}
\end{figure}

We have discussed our multi-resolution encoder with multiple block sizes to adapt to the point cloud structure. However, with smaller block sizes, an implicit context model (using the content of the block) will be less efficient because the context may be too small. Therefore, we extend the context of each block to the encoded voxels that are above and on the left of the current voxel (causality constraint). Figure \ref{fig:context extension} illustrates the context before and after extension. Before extending the context, to encode voxel $v_c$, only voxels from $v_1$ to $v_{i-1}$ in Figure \ref{fig:context extension}(a)  are considered as contexts. After extending the context to the bigger block, the context is now composed of all voxels in the blue area in Figure \ref{fig:context extension}(b). The white area represent inactive voxels, i.e., not used in Eq.~\eqref{eq:p(v)}. Extending the context does not change the partitioning algorithm discussed above, although it might change the optimal selected partitions. Also, the causality is still enforced as long as we use masked filters in our network. 

However, extending to a larger context is not always efficient when the extension area is sparse or contains noise, therefore we employ a rate-optimized block extension decision. To limit the computational complexity, we only allow certain combinations of block sizes and extension sizes, as shown in  Table \ref{table:extending block size}. To encode a block with context extension, in Algorithm \ref{algo:proposed_method}, we encode a block with all the possible extension sizes and select the best one in terms of bpov. In total, we build 5 models for 5 input sizes which are $\{ 128,64,32,16,8\}$ in the context extension mode.

\newcolumntype{L}[1]{>{\centering\arraybackslash}p{#1}}

\begin{table}[b]
\caption{Extending block size}
\centering
\begin{tabular}{M{2cm}|L{3cm}}
\hline
\begin{bf} Block size \end{bf}&\begin{bf} Extending block size \end{bf}  \\
\hline
 64 & 128,64\\
 32 & 64,32\\
 16 & 64,32,16\\
 8 & 64,32,16,8\\
\hline
\end{tabular}
\label{table:extending block size}
\end{table}

\subsection{Data augmentation} 
\begin{figure}[tb]
\captionsetup{justification=raggedright}
%\captionsetup{singlelinecheck = false, format= hang, justification=raggedright, font=small, labelsep=space}
\centering
\includegraphics[width=0.9\linewidth]{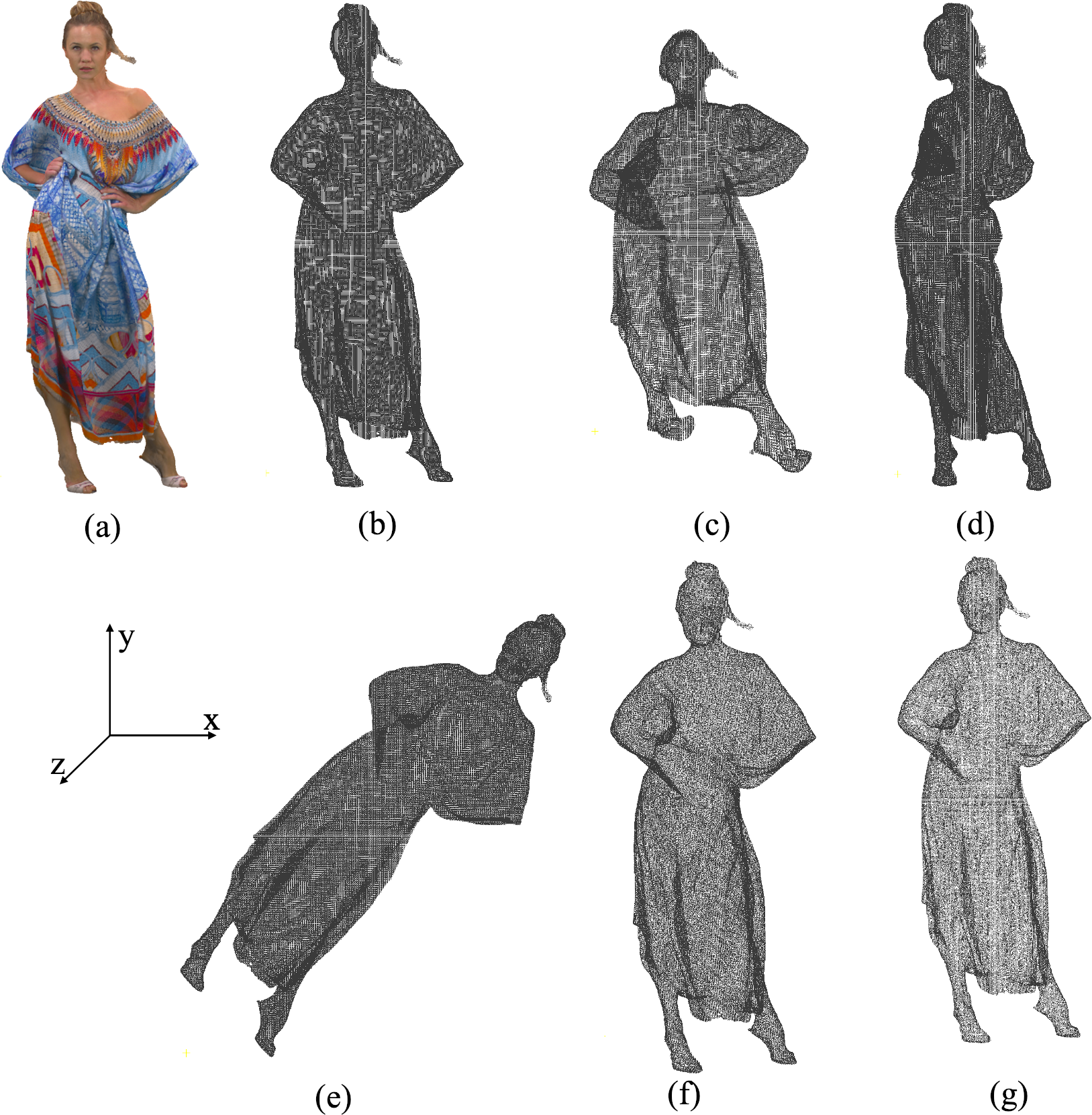}
\caption{Example of data augmentation applied on the Longdress point cloud. (a) Original; (b) After removing color attributes; (c),(d),(e) Rotation with $\theta=45^{\circ}$ on $x,y$ and $z$ axis; (f),(g) Sampling rate  $f_{s}=0.7$ and $f_{s}=0.4$ }
\label{fig:dataaugment}
\end{figure}

In order to train more robust probability estimation models and to increase the generalization capabilities of our model, we employ data augmentation techniques specifically suited for PCC.
In particular, we observed that methods based on convolutional neural networks are especially sensitive to changes in PC density and acquisition noise. Therefore, in addition to typical rotation and shifting data augmentation used for other PC analysis tasks \cite{alonso20203dmininet, wang2020multiscale}, we also consider here alternative techniques, such as downsampling. Note that even though our VoxelDNN operates on voxel domain, to reduce the complexity, all input pipelines process point clouds in the form of $x,y,z$ coordinates before converting into dense block in the final step.
Specifically, for each generated block from the training datasets, we rotate them by an angle $\theta$ around each $x,y,z$ axis. In addition, to adapt to varying density levels of the test point clouds, we randomly remove points from the original block as well as rotated blocks with the sampling rate $f_{s}$ ($f_s \in \left[ 0, 1\right]$) over the total points. Figure \ref{fig:dataaugment} shows our data augmentation methods applying on Longdress point cloud from MPEG. \\

\section{Experimental Results}
\label{performanceeval}
\subsection{Experimental Setup}
\begin{table}[t]
\caption{Training and Testing Point Clouds}
\centering
\resizebox{0.99\linewidth}{!}{
\begin{tabular}{|l|c|c|c|l|c|}
\cline{1-3}\cline{5-6}
\multicolumn{3}{|c|}{Training Set}&&\multicolumn{2}{c|}{Test Set}\\
\cline{1-3}\cline{5-6}
\multicolumn{1}{|c|}{Point Cloud}&$\#$ Fr& $\rho$&&\multicolumn{1}{c|}{Point Cloud}& $\rho$\\
\cline{1-3}\cline{5-6}
\multicolumn{3}{|c|}{\textit{MVUB, 10 bits}} & &\multicolumn{2}{c|}{\textit{MVUB, 10 bits, dynamic upper body}} \\
Andrews&6&1.70  &&Phil&1.64\\
David&5&1.65  && Ricardo&1.77\\
Sarah&6&1.72  && &\\
&&&& \multicolumn{2}{c|}{\textit{8i, 10 bits, dynamic full body}} \\
\multicolumn{3}{|c|}{\textit{8i, 10 bits}}&&Redandblack&1.49\\
Soldier&9&1.51 &&Loot&1.43\\
Longdress&9&1.52 &&Thaidancer&1.68\\
&&&&Boxer&1.56\\
\multicolumn{3}{|c|}{\textit{CAT1, 10 bits}}&& &\\
Facade&1&1.20 &&\multicolumn{2}{c|}{\textit{CAT1, 10 bits, static cultural heritage}} \\
Egyptian mask&1&0.12   &&Frog&1.13\\
Statue klimt&1&0.89  && Arco Valentino&0.45\\
Head&1&1.43  &&Shiva&0.88\\
House w/o roof&1&1.21  &&&\\
&&&&\multicolumn{2}{c|}{\textit{USP, 10 bits, static cultural heritage}}\\
\multicolumn{3}{|c|}{\textit{ModelNet40, 9 bits}} && BumbaMeuBoi&0.18\\
200 largest PCs&200& 1.53 && RomanOilLight&0.94\\
\cline{1-3}\cline{5-6}
\cline{1-3}\cline{5-6}
\end{tabular}}
\label{table:traintestpcs}
\end{table}

\subsubsection{Training dataset} We consider point clouds from different and varied datasets, including ModelNet40 \cite{wu20153d} which contains 12,311 models from 40 categories and three smaller datasets: MVUB \cite{loop2016microsoft}, MPEG CAT1 \cite{noauthor_common_nodate} and 8i \cite{d20178i,8i}. We uniformly sample points from the mesh models from ModelNet40 and then scale them to voxelized point clouds with 9 bit precision. To enforce the fairness between the smaller datasets in which we select point clouds for testing, point clouds from MPEG CAT1 are sampled to 10 bit precision as in MVUB and 8i. In addition, we measure the \textit{local density} $\rho$ of a point cloud, computed as the average portion of occupied voxels in the blocks of size 64, that is: 
\begin{equation}
    \rho = \frac{1}{N_{\mathcal{B}}}\times\sum_{\mathcal{B}_i \in \mathcal{B}} \frac{100 \times \text{number of points in }\mathcal{B}_i}{64^3}\ \ (\%)
    \label{rhodensity}
\end{equation}
%(x_{max}-x_{min})(y_{max}-y_{min})(z_{max}-z_{min})
where $\mathcal{B}$ is the set of occupied blocks of size $64$, and $N_{\mathcal{B}}$ is the cardinality of $\mathcal{B}$. 
% We found that $d=64$ provides a representative $\rho$ for the point cloud density.  
The higher the value of $\rho$ is, the denser the point cloud. The selected point clouds, number of frames as well as $\rho$ of the training data are shown in Table \ref{table:traintestpcs}. 
\par To train a VoxelDNN model of size $d$ we divide all selected PCs into occupied blocks of size $d\times d\times d$. Table \ref{table:noblocks} reports the number of blocks from each dataset for training, with the majority coming from the ModelNet40 dataset. For the models trained with data augmentation, we apply rotation with $\theta=45^\circ$ on $x,y,z$ axis and then sampling from all blocks with sampling rate $f_s=[0.7;0.4]$. In total, we augment from each block to 12 variations in terms of density and rotation which significantly increase the volume and diversity of our training set.
%Smallest bounding box removes most of the empty space of each occupied blocks in $B$. And we aim to measure local density of point clouds, therefore, in Equation \ref{rhodensity}, number of occupied voxels is normalized by Smallest bounding box instead of the original block.
\begin{table}[t]
\caption{Number of blocks in the training sets of each model.}
\centering
\resizebox{0.97\linewidth}{!}{ \begin{tabular}{lR{0.8cm}R{0.8cm}R{0.8cm}R{1.2cm}R{1cm}}
\hline
\begin{bf}  \end{bf}
&\begin{bf}MVUB\end{bf}
& \begin{bf}8i\end{bf} 
&\begin{bf}CAT1\end{bf}
& \begin{bf}ModelNet40\end{bf} 
& \begin{bf}Total\end{bf} \\
\hline
Model 128& 1516&1101 &677 & 2860&6154\\
Model 64& 5777&4797 &2777 & 11147&24498\\
Model 32& 22082&20436 &15243 & 50611&108372\\
Model 16& 87578&86106 &45626 & 224951&444261\\
Model 8& 354617&349760 &180037 & 986253&1870667\\
\hline
\end{tabular}}
\label{table:noblocks}
\end{table}

\subsubsection{Test data} We evaluate the performance of our approach on a diverse set of point clouds in terms of spatial density and content type. All selected point clouds are either used in MPEG Common Test Condition or JPEG Pleno Common Test Condition to evaluate point cloud compression methods. As shown in Table \ref{table:traintestpcs} the test PCs can be categorized into four sets:
\begin{itemize}
    \item \textbf{MVUB}: Microsoft Voxelized Upper Bodies \cite{loop2016microsoft} - a dynamic voxelized point cloud dataset containing five subjects. For testing, we randomly select 2 frames from \textit{Phil} (frame number 10) and \textit{Ricardo} (76) sequences which are both very dense (high $\rho$) with smooth surfaces.
    \item \textbf{8i}: Dense point clouds from 8i Labs. They are also dynamic voxelized point clouds but each sequence contains the full body of a human subject. In the test set, \textit{loot} (1000) and \textit{redandblack} (1510) are from 8i Voxelized Full Bodies (8iVFB v2) \cite{d20178i} while \textit{boxer} and \textit{thaidancer} are selected and downsampled to 10 bits from 8i Voxelized Surface Light Field (8iVSLF) dataset \cite{8i}.
    \item \textbf{CAT1}: static point clouds for cultural heritage and other related 3D photography applications \cite{noauthor_common_nodate}. We select \textit{Arco\_Valentino\_Dense\_vox12, Frog\_00067\_vox12, and Shiva\_00035\_vox12} from this dataset and downsample to 10 bits to validate the performance of our method. PCs from this dataset are less dense compared to the previous two datasets. \textit{Frog\_00067} has smoother surfaces compared to the other two PCs which contain rough surfaces.
    \item \textbf{USP}: an inanimate dataset from the University of S\~ao Paulo, Brazil, related to cultural heritage with 10 bits geometry precision \cite{usp}. \textit{BumbaMeuBoi} and \textit{RomanOilLight} are two selected point clouds from this dataset. PCs from USP dataset have simple shape with smooth surfaces. \textit{BumbaMeuBoi} is the sparsest PC in our test set with the smallest $\rho$. 
\end{itemize}
\par
\begin{figure}[tb]
%\captionsetup{singlelinecheck = false, format= hang, justification=raggedright, font=small, labelsep=space}
\captionsetup{justification=raggedright}
\centering
\includegraphics[width=0.99\linewidth]{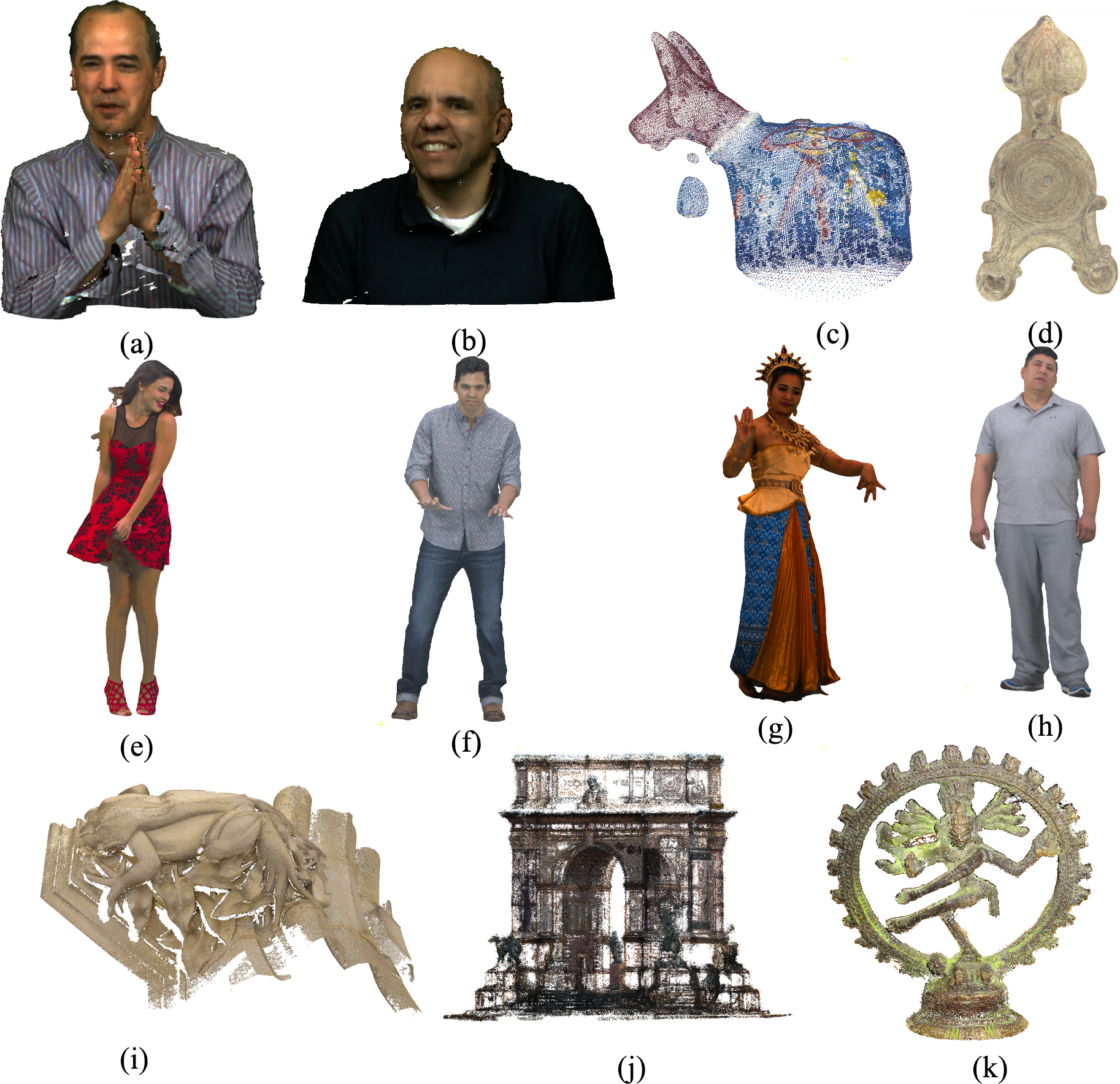}
\caption{Point clouds in the test set. (a) Phil, (b) Ricardo (c) BumbaMeuBoi (d) RomanOilLight, (e) Redandblack, (f) Loot, (g) Thaidancer (h) Boxer, (i) Frog, (j) Arco Valentino, (k) Shiva.}
\label{fig:test pcs}
\end{figure}

 Figure \ref{fig:test pcs} illustrates the test point clouds.

\subsubsection{Training procedure} 
We train 5 models for 5 input block sizes, i.e., 128, 64, 32, 16, 8. The mini-batch sizes are 1, 8, 64, 128, 128, respectively. Our models are implemented in TensorFlow and trained with Adam \cite{adam} optimizer, a learning rate of 0.001 for 80 epochs on a GeForce RTX 2080 GPU.\footnote{The source code, as well as the trained models, will be made publicly available upon acceptance of the paper.}
% \subsubsection{Evaluation procedure}

\subsection{Performance evaluation and ablation studies}
In the following, we evaluate the performance of the proposed approach as well as the impact of its various components. We start with models without data augmentation nor context extension in order to study the optimal maximal partitioning depth for our method and establish a baseline for the evaluation. 
Next, on top of the best encoder in this experiment (\textbf{Baseline}), we separately add data augmentation (\textbf{Baseline~+~DA}) and context extension (\textbf{Baseline~+~CE}). Finally, \textbf{Baseline~+~DA~+~CE} incorporates both data augmentation and context extension. We compare our method against the state-of-the-art point cloud compression method G-PCC from MPEG \cite{graziosi2020overview} which has a dedicated lossless geometry mode for static point clouds. We report the number of bits per occupied voxel (bpov) for each test point cloud and the average per dataset.

\setlength{\textfloatsep}{20pt}

In all experiments, the high-level octree plus partitioning signal are directly converted to bytes without any compression. For the encoders with context extension, we signal the selected size using two bits (maximum 4 options on block 8). This information is also directly converted to bytes in the bitstream. On average, signaling bits account for $2.44\%$ of the bitstream.

\setlength{\textfloatsep}{20pt}
\begin{center}
\centering
\begin{table*}[ht]
\caption{Average rate in bpov per dataset at different partitioning levels and the gain over the encoder with 1 partitioning level.}
\centering
\resizebox{0.7\linewidth}{!}{

\begin{tabular}{|P{0.9cm}|l|R{0.9cm}|R{0.9cm}|R{1cm}|R{0.9cm}|R{0.99cm}|R{0.9cm}|R{0.99cm}|}
\cline{3-9}
% \multicolumn{2}{|c||}{\begin{bf} Test PC \end{bf}}
\multicolumn{2}{c|}{}
& \multicolumn{1}{c|}{\begin{bf} 1 level \end{bf}}
& \multicolumn{2}{c|}{\begin{bf} 2 levels \end{bf}}
& \multicolumn{2}{c|}{\begin{bf} 3 levels \end{bf}}
& \multicolumn{2}{c|}{\begin{bf} 4 levels \end{bf}}
\\
\hline
Dataset&Point Cloud&bpov &bpov&Gain &bpov&Gain&bpov&Gain\\
\hline
\multirow{3}{*}{MVUB}&Phil  &0.8943  &0.8295 &-7.25\% &0.8206 &-8.24\% &  0.8205&-8.25\%\\
\cline{2-9}
&Ricardo &0.8109  &0.7511 &-7.37\% &0.7440 &-8.25\% &  0.7440&-8.25\%\\
\cline{2-9}
&\textbf{Average} &\textbf{0.8256 } &\textbf{0.7903 }&\textbf{-7.31\%} &\textbf{0.7823 }&\textbf{-8.25\%} & \textbf{ 0.7823}&\textbf{-8.25\%}\\
\hline
\multirow{5}{*}{8i} &Redandblack & 0.7920&0.7269 & -8,22\%& 0.7191& -9.20\%& 0.7190&-9.22\% \\
\cline{2-9}
 &Loot& 0.7017&0.6347 & -9.56\%& 0.6271& -10.63\%& 0.6271&-10.63\% \\
\cline{2-9}
&Thaidancer & 0.7941&0.7360 & -7.32\%& 0.7298& -8.10\%& 0.7297&-8.11\% \\
\cline{2-9}
&Boxer&0.6462&0.5960 & -7.77\%& 0.5901& -8.68\%& 0.5900&-8.70\% \\
\cline{2-9}
&\textbf{Average}&\textbf{ 0.7335}&\textbf{0.6734} &\textbf{ -8.22\%}&\textbf{ 0.6665}&\textbf{ -9.15\%}&\textbf{ 0.6665}&\textbf{-9.16\%} \\
\hline
\multirow{4}{*}{CAT1} &Frog& 1.9497 &1.8406 & -5.60\%&1.8216 & -6.57\%&  1.8214&-6.58\% \\
\cline{2-9}
 &Arco Valentino& 5.4984 &5.2947 & -4.52\%&5.2051 & -5.33\%&  5.2050&-5.34\% \\
\cline{2-9}
 &Shiva &3.7964 &3.6632 & -3.51\%&3.6400 & -4.01\%&  3.6403&-4.11\% \\
\cline{2-9}
  &\textbf{Average}&\textbf{ 3.7482} &\textbf{3.5845} &\textbf{ -4.54\%}&\textbf{3.5569 }& \textbf{-5.31\%}&\textbf{  3.5556}&\textbf{-5.34\% }\\
  \hline
\multirow{3}{*}{USP} &BumbaMeuBoi& 6.3618 &  5.8235&-8.46\% &5.7305&-9.92\% & 5.7305&-9.92\% \\
\cline{2-9}
&RomanOilLight&  1.8708 &1.7157 & -5.14\%&1.7030 & -5.84\%&  1.7030&-5.84\% \\
 \cline{2-9}

&\textbf{Average}&\textbf{ 4.0853} &\textbf{3.7696} & \textbf{-6.80\%}&\textbf{3.7168 }& \textbf{-7.88\%}& \textbf{ 3.7168}&\textbf{-7.88\%} \\
\hline
\hline
\end{tabular}}
\label{table:increase partitioning level}
\end{table*}
\end{center}

\subsubsection{Optimal maximum partition depth} 
To evaluate the effectiveness of the partitioning scheme, we increase the maximum partitioning level from 1 to 5, corresponding to a minimum block size of  64, 32, 16, 8, and 4. 
% As explained in Subsection \ref{ssec:multires}, we could use a bigger model size to predict the distribution of smaller block. 
As 3D convolution is not able to efficiently exploit voxel relations on a very small receptive field, we do not train a  separate model for block 4.  Instead, we use the model trained on blocks of size 8 to predict its probabilities. 

\begin{figure*}[tb]
\captionsetup{justification=raggedright}
\centering
\includegraphics[width=0.87\linewidth]{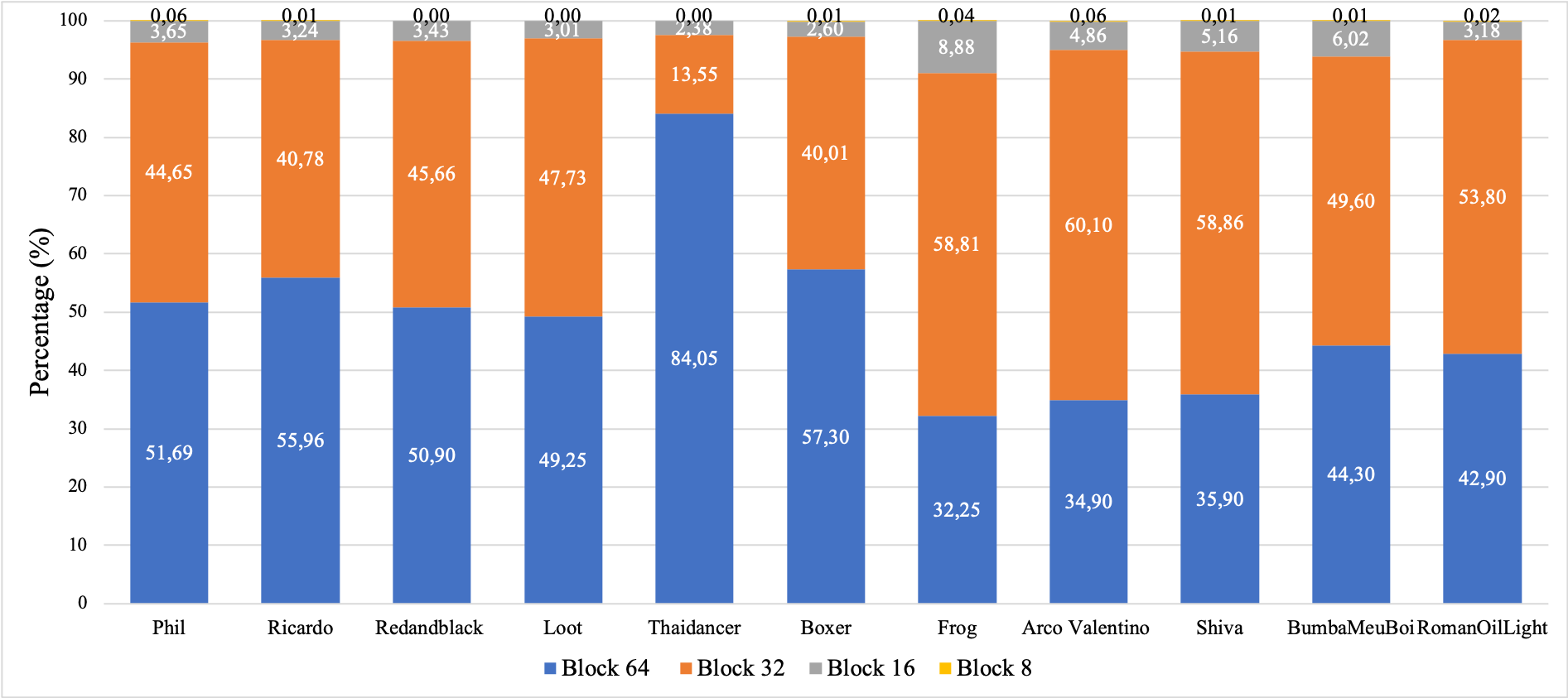}
\vspace{0.1cm}
\caption{Percentage of occupied voxels encoded in each partition size. From top to bottom: block 8, 16, 32, 64. Most of occupied voxel are encoded in block 64 and block 32. }
\label{fig:ocv_per_block}
\end{figure*}
%\vspace{15mm}
\setlength{\textfloatsep}{20pt}

Table \ref{table:increase partitioning level} shows the average bpov of our encoder on the 4 test datasets at 4 partitioning levels. The results with 5 partitioning levels are identical to 4 partitioning levels and are not shown in the table. We observe that, as partitioning levels increases, the corresponding gain over single-level also increases. However, adding the $3^{rd}$ and $4^{th}$ level yields only a slight improvement compared to adding the $2^{nd}$ level. This can be explained with Figure \ref{fig:ocv_per_block} showing the percentages of occupied voxels in each partition size. We  observe that most  voxels are encoded using blocks 64 and 32, while very few voxels are encoded using blocks of smaller size. Adding more partitioning levels enables to better adapt to point cloud geometry, however, this is not often compensated by a bitrate reduction of the sub-blocks, since in the smaller partitions the encoder has access to limited contexts, resulting in less accurate probability estimation. However, there is an increase in the portion of block 32 and 16 on CAT1 and USP compared to MVUB and 8i. This reflects the density characteristics of each dataset: on sparser datasets (CAT1 and USP), the algorithm tends to partition point cloud into smaller blocks to eliminate as much empty space as possible. Based on these observations, we use a maximum of 4 partitioning levels for our baseline codec in later experiments.
%\input{partitioning-result-per_pc}
%It can be seen from $\rho$ value for each point cloud from Table \ref{table:traintestpcs} that PCs from MVUB and 8i datasets are denser compare to PCs from CAT1 and USP.

\begin{center}
\centering
\begin{table*}[ht]
\caption{Average rate in bpov of proposed method and percentage gains compared with MPEG G-PCC (negative percentages mean bitrate reduction).}
\resizebox{0.99\linewidth}{!}{ \begin{tabular}{|P{0.80cm}|l|R{0.9cm}||R{0.9cm}|R{1.2cm}|R{0.9cm}|R{1.2cm}|R{0.9cm}|R{1.2cm}|R{0.9cm}|R{1.2cm}|}
\cline{3-11}
% \multicolumn{2}{|c||}{\begin{bf} Test PC \end{bf}}
\multicolumn{2}{c|}{}
& \begin{bf}G-PCC\end{bf}
& \multicolumn{2}{c|}{\begin{bf} Baseline \end{bf}}
& \multicolumn{2}{c|}{\begin{bf} Baseline + DA \end{bf}}
& \multicolumn{2}{c|}{\begin{bf} Baseline + CE \end{bf}}
& \multicolumn{2}{c|}{\begin{bf} Baseline + DA + CE\end{bf}}
\\
\hline
Dataset&Point Cloud&bpov&bpov&Gain over G-PCC &bpov&Gain over G-PCC &bpov&Gain over G-PCC&bpov&Gain over G-PCC\\
\hline

\multirow{3}{*}{MVUB}&Phil & 1.1617 &0.8205 &-29.37\% &0.8954 & -22.92\%&0.7601 & -34.57\%&  0.8252&-28.97\% \\ 
\cline{2-11}
&Ricardo & 1.0672 & 0.7440&-30.28\% &  0.8235&-22.84\% &0.6874&-35.59\% &  0.7572&-29.05\% \\
\cline{2-11}
&\textbf{Average} &\textbf{1.1145}  &\textbf{0.7823}&\textbf{-29.83\%} &\textbf{0.8595 }&\textbf{-22.88\%} &\textbf{07238. }&\textbf{-35.06\%}& \textbf{0.7912}&\textbf{-29.01\% } \\
\cline{2-11}
\hline

\multirow{5}{*}{8i}&Redandblack &1.0899  & 0.7190&-34.3\% &0.7772 &-28.69\% & 0.6645&-39.03\%& 0.7003&-35.75\% \\
\cline{2-11}
&Loot & 0.9524 &0.6271 & -34.16\%&0.6282 &-34.04\% &0.5766 &-39.46 \% &0.6084 &-36.12\%\\
\cline{2-11}
&Thaidancer &0.9985 &0.7297 &-26.92\% &0.7253 &-27.36\% &0.6769 &-32.21\%&0.6627   &-33.63\%\\
\cline{2-11}
&Boxer&0.9479  & 0.5900&-37.76\% &0.6573 &-30.66\%&0.5503 &-41.95\% &0.5906 &-37.69\% \\
\cline{2-11}
&\textbf{Average} &\textbf{0.9972} &\textbf{0.6665} &\textbf{-33.22\%} &\textbf{0.6870} &\textbf{-30.19\% }&\textbf{0.6171} &\textbf{-38.12\% }& \textbf{ 0.6405}&\textbf{-35.77\% }\\
\hline

\multirow{4}{*}{CAT1}&Frog&1.9085   & 1.8214&-4.56\% &1.7662 &-7.64\% & 1.6971&-11.08\%& 1.7071&-10.55\% \\
\cline{2-11}
&Arco Valentino & 4.8119 & 5.2050&+8.17\% &5.0639 &+5.24\% & 4.9923&+3.75\%& 4,9900&+3.70\% \\
\cline{2-11}
&Shiva&3.6721 & 3.6403&-0.87\% &3.5838 &-2.04\% & 3.4619&-5.72\%& 3.5135&-4.32\% \\
\cline{2-11}
&\textbf{Average} &\textbf{3.4642} &\textbf{3.5556} &\textbf{+0.91\%} &\textbf{3.7413} &\textbf{-1.54\% }&\textbf{3.3838} &\textbf{-2.32\% }& \textbf{3.4035 }&\textbf{-3.72\% }\\
\hline

\multirow{3}{*}{USP}&BumbaMeuBoi & 5.4522   & 5.7305&+5.10\% &5.3831 &-1.27\% & 5.3580&-1.73\%& 5.066&-7.08\% \\
\cline{2-11}
&RomanOiLight & 1.8604  & 1.7030&-8.46\% &1.7319 &-6.91\% & 1.6130&-13.30\%& 1.6231&-12.76\% \\
\cline{2-11}
&\textbf{Average}  &\textbf{3.6563}  &\textbf{3.7168}&\textbf{-1.68\%} &\textbf{3.5575}&\textbf{-4.09\%} &\textbf{3.4855 }&\textbf{-7.51\%}& \textbf{3.4855}&\textbf{-9.91\% } \\
\cline{2-11}
\hline
\hline
\end{tabular}}
\label{table:result table}
\end{table*}
\end{center}

\subsubsection{Comparison with G-PCC} 
In table \ref{table:result table}, we report the output bitrate of our methods to compare with MPEG G-PCC. Both our method and G-PCC perform better on dense PCs while having higher rates on sparser PCs. 
%Our \textit{Baseline} is the encoder with 4 partitioning level.
Compared to G-PCC, the \textbf{Baseline} encoder obtains a significant gain -- over $29\%$ bitrate reduction on dense point clouds from MVUB and 8i dataset. On CAT1 and USP datasets, our method achieves a comparable rate with G-PCC. In particular, for Arco Valentino and BumbaMeuBoi, the two point clouds having the lowest $\rho$, our baseline codec yields a rate higher than G-PCC ($+8.17\%$ and $+5.10\%$, respectively). For point clouds with high local density levels, our VoxelDNN could efficiently leverage the relations between voxels and predict more accurate probability. In contrast, probability prediction is less accurate on sparser point clouds. 

This can be partially solved by adding data augmentation during training.
Indeed, by random subsampling the point clouds in the training set, VoxelDNN learns to predict more accurate probabilities when the point cloud is less dense. \textbf{Baseline + DA} yields higher gains over G-PCC on CAT1 and USP compared to \textbf{Baseline}, with average bitrate reductions of about $1.54\%$ and $4.09\%$, respectively. On the other hand, we observe a small degradation of the performance compared to \textbf{Baseline} for denser datasets, such as MVUB and 8i dataset. This is somehow expected, as data augmentation increases the generalization capability of VoxelDNN, which instead is more adapted to denser PCs in the baseline mode.
\par The encoder with context extension, \textbf{Baseline + CE}, obtains a better rate on all test point clouds compared to the  \textbf{Baseline} encoder, regardless of the density, with an average further bitrate reduction of $4.8\%$ over G-PCC. The cost to be paid for this performance improvement is a higher computational complexity in the encoding process.
%Adaptively expanding the context to neighboring blocks has clearly shown the effectiveness, however, in terms of complexity, this method increases the complexity when trying all possible expanding options for each block.

\par The last two columns of Table \ref{table:result table} show the experiment results for the encoder incorporating both data augmentation and context extension, \textbf{Baseline + DA + CE}. On average, we have a higher gain than \textbf{Baseline} and \textbf{Baseline + DA} because of the Context Extension. As expected, comparing with \textbf{Baseline + CE}, \textbf{Baseline + DA + CE} has increasing gains on CAT1 and USP datasets while obtaining a lower gain on MVUB and 8i datasets. Despite the different performance trends for different densities of the input point clouds, we obtain, on average, a bitrate reduction of $20.17\%$ compared to G-PCC. Note that, in practice, if the characteristics of point cloud to be coded are known in advance, our approach is flexible, in that we could deploy different models targeting a specific application (cultural heritage, tele-immersive conferencing, etc.) and content type to obtain the best compression rate.

\begin{figure*}[tb]
%\captionsetup{singlelinecheck = false, format= hang, justification=raggedright, font=small, labelsep=space}
\captionsetup{justification=raggedright}
\centering
\includegraphics[width=0.75\linewidth]{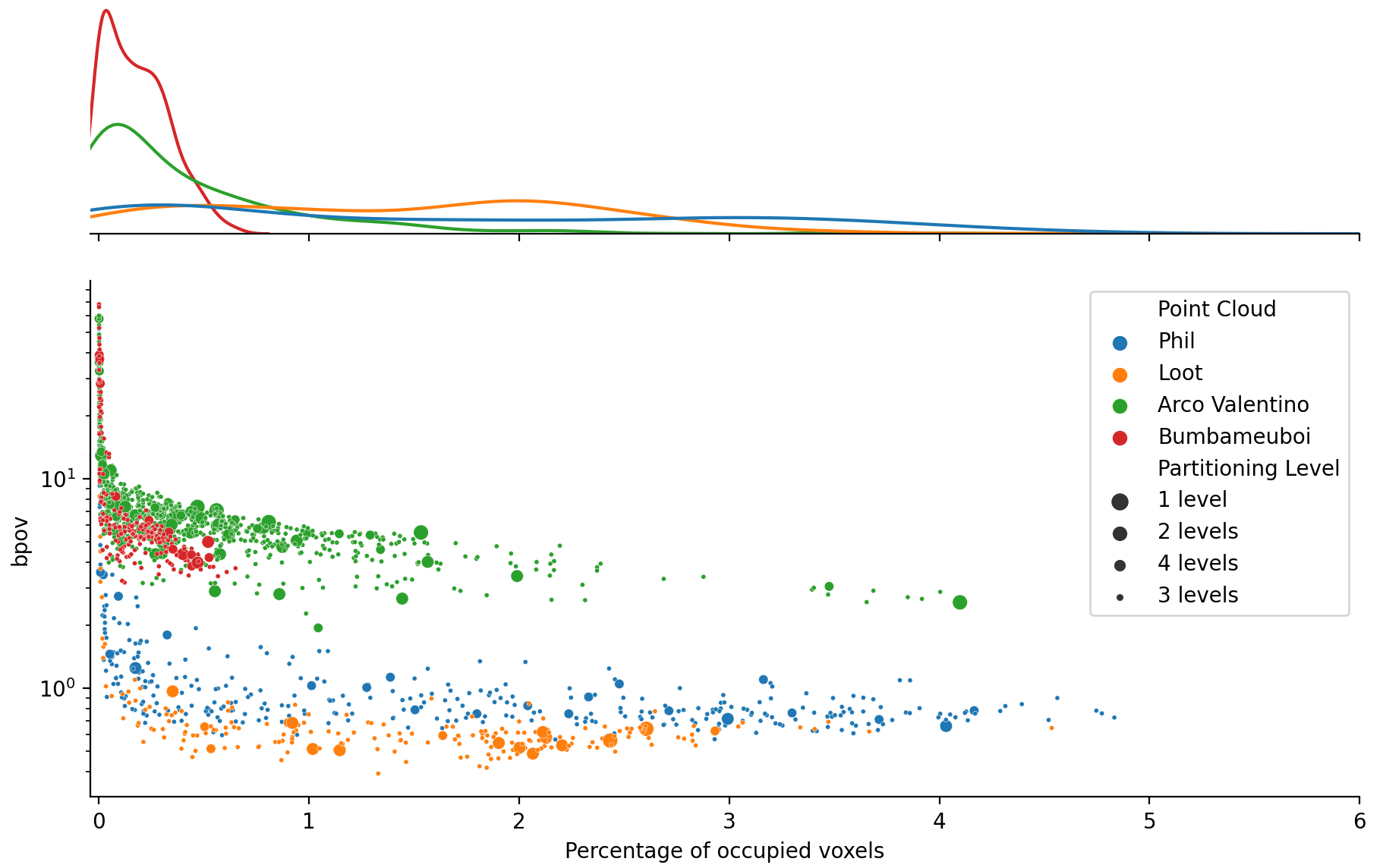}
\caption{Performance on block 64 on four test point clouds. Each point corresponds to a block 64 with percentage of occupied voxels ($\rho$) and bpov ($\log$ scale) performance of that block. The size of each point indicates the partitioning level and each partitioning level was sized according to its frequency. Higher points indicate that VoxelDNN is performing worse. The marginal distributions of occupied voxels for each point cloud are on the top of the scatter plot. }
\label{fig:bpovperformace}
\end{figure*}
%Steeper slopes indicate that VoxelDNN is performing worse.
%
\begin{figure}[tb]
%\captionsetup{singlelinecheck = false, format= hang, justification=raggedright, font=small, labelsep=space}
\captionsetup{justification=raggedright}
\centering
\includegraphics[width=0.99\linewidth]{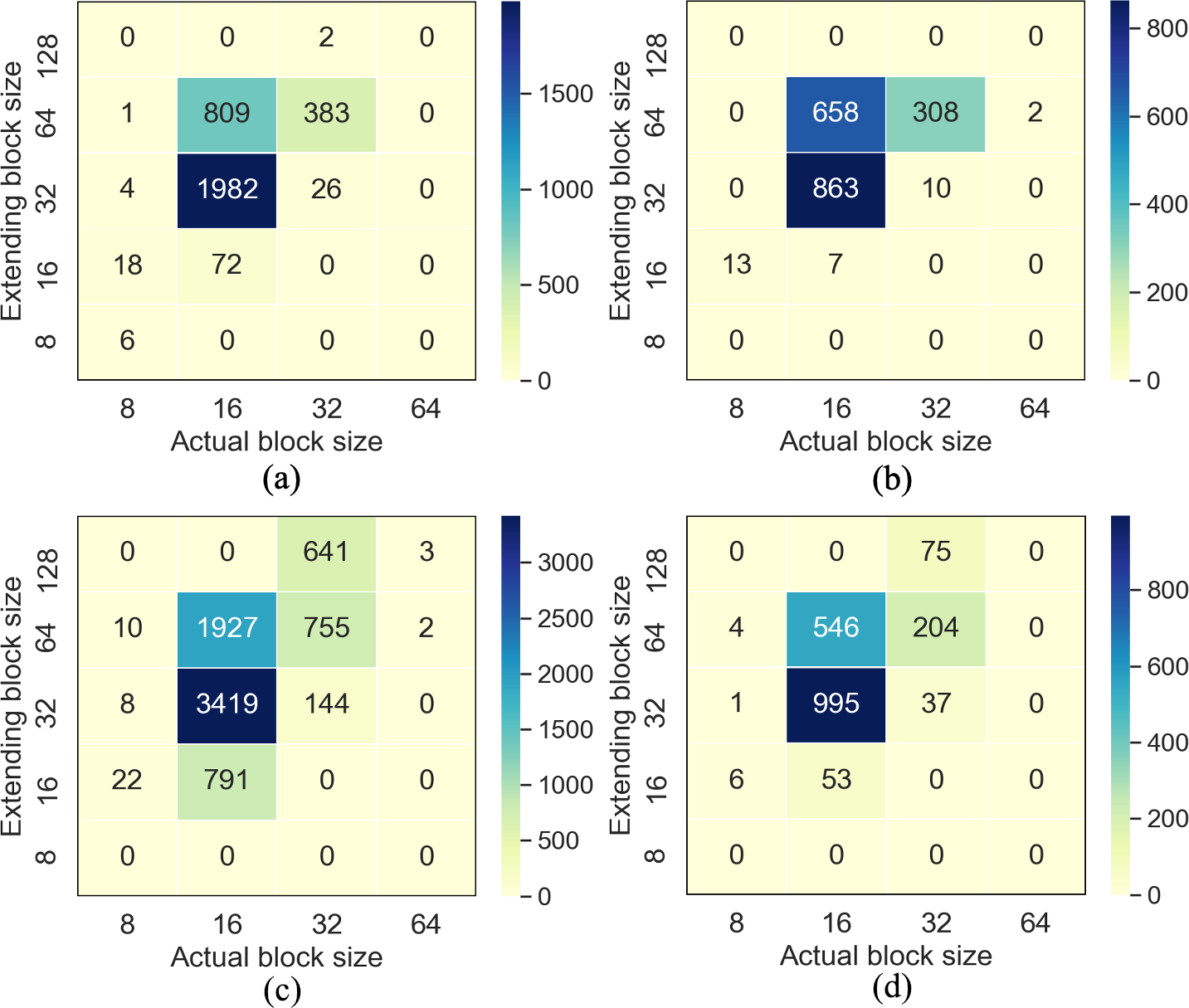}
\caption{Number of extending block size for each block. (a) Phil, (b) Loot, (c) Arco Valentino, (d) BumbaMeuBoi. Most of the time, the encoder extend the context to neighboring voxels instead of independently encoding a block. }
\label{fig:extendingblocksize}
\end{figure}
\setlength{\textfloatsep}{20pt}

\subsubsection{Effect of PC content and density on coding performance} 

%\textcolor{red}{    The text of this section should be modified and shortened to make the message more clear. From the current version and the figure, my understanding is that the subsection is about the effect of density on coding.  It is not clear from the figure why different contents have different slopes. If the coding performance depends on the local density, then shouldn't all the contents lie on the same line? in other words, the fact that different contents have different slopes cannot be explained by the local density only. For ex, you can take a block in Look with 4000 occupied voxels, and a block in Arco Valentino with 4000 occupied voxels, and they require a totally different number of bits... can you explain this better?}

In order to better understand the performance of our codec for different types of content, we plot in Figure~\ref{fig:bpovperformace} the average bpov as a function of the percentage of occupied voxels for each block 64 of \textit{Phil, Loot, Arco Valentino} and \textit{BumbaMeuBoi} with the \textbf{Baseline + DA + CE} encoder. Notice that each block 64 can be split up to different partition levels, indicated by the size of the dots in the figure. The distribution of the density of blocks 64 is shown in the top panel. 

From this figure, we can draw some observations.
First, most of blocks are partitioned into 3 levels (smallest dots) and the majority of the remaining blocks are partitioned into 2 or 4 levels. Second, in each point cloud, denser blocks are easier to compress, as mentioned before, due to the better capabilities of convolution to capture spatial relations. On the other hand, our approach becomes inefficient when the blocks are less dense, and the bitrate associated to the very sparse blocks rapidly grows by an order of magnitude compared to the rest. This phenomenon is true for all kinds of contents, although it has a stronger effect when the block density distribution is skewed to the left, such as for \textit{Arco Valentino} or \textit{BumbaMeuBoi}, which have the highest bitrates in our experiments. 
%This can be explained by the fact that dense blocks provide better context to VoxelDNN. Moreover, the sparse blocks must pay a fixed overhead-bits for partitioning signal while contain just few occupied voxels.
\par
\begin{figure*}[tb]
%\captionsetup{singlelinecheck = false, format= hang, justification=raggedright, font=small, labelsep=space}
\captionsetup{justification=raggedright}
\centering
\includegraphics[width=0.95\linewidth]{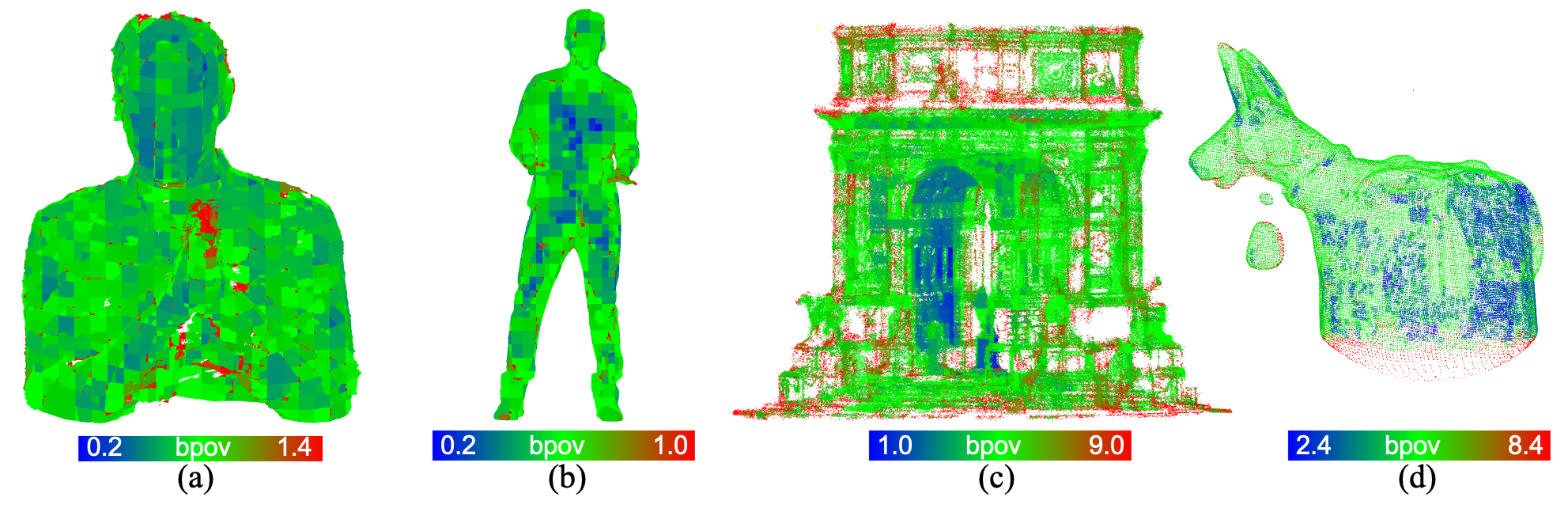}
\caption{Output geometry bitrate in bpov per block. (a) Phil, (b) Loot, (c) Arco Valentino, (d) BumbaMeuBoi. The heatmap bar below each subfigure shows the minimum and maximum bpov and the corresponding color. }
\label{fig:bitallocation}
\end{figure*}

 We can also observe a content-dependence trend in the figure, which appears like a vertical offset for different PCs. \textit{Arco Valentino} and \textit{RomanOilLight} overall have higher bpov compared to \textit{Phil} and \textit{Loot} with the same number of occupied voxels. This suggests that local density alone is not the only factor affecting the performance of our approach, but that somehow higher-order statistics enter into play. We will speculate more about this behaviour when discussing the bitrate allocation in Figure~\ref{fig:bitallocation}. Further analysis of this trend, as well as how to take better into account the PC characteristics to improve coding performance, are left to future work.
% Given the same number of occupied voxels, those voxels can be spread over the whole block or landed on rough/smooth surfaces in which defenitely influence the performance. Therefore, the local density characteristic is not the only factor that affect the performance but also the content itself. We will further investigate this with Figure \ref{fig:bitallocation}. 

\vspace{15mm}
%\setlength{\textfloatsep}{15pt}% Remove

%\par First, we observe that most of blocks are partitioned into 3 levels (smallest points) and the majority of the remaning blocks are partitioned into 2 or 4 levels. However, in each point cloud, most of blocks are lying on the same slope and thus, there is a linear relation between the number of occupied voxels and encoded bits of block 64. We found similar characteristics on other point clouds. Those evidences show that the partitioning efficiently remove the redundancies within blocks and the output bits on each each block is significantly influenced by the occupied voxels even when most of voxels are empty. 
%\par However, different point clouds give different slopes and a higher slope is, a better VoxelDNN performance. Therefore, the local density characteristic is not the only factor that affect the performance but also the content itself. We will further investigate this with Figure \ref{fig:bitallocation}.
%\\

\subsubsection{Selection of context extension and impact on the partitioning}
Figure \ref{fig:extendingblocksize} shows how many times an extended block size is selected in the \textbf{Baseline + DA + CE} experiments. First, it can be seen that in most  cases our encoder choose to extend the context to encode the current block, and mostly the immediate larger size is selected. By extending context to exploit geometry information from the neighboring voxels, VoxelDNN can leverage a larger amount of information and predict a better probability. In most  cases where the encoder does not extend the context, the blocks are on the border of the volume, corresponding to a mostly empty  extending area.

By summing the quantities in each column, we obtain the number of blocks which are encoded using each block size and we observe that large parts of the point cloud are partitioned into block 32 or 16. This is in contrast with the previous observation on baseline experiments where the most frequent partitions are 64 and 32 (Figure~\ref{fig:ocv_per_block}). This has an intuitive explanation: without context extension, small block sizes of 32 or 16 were insufficient to provide a representative enough context for VoxelDNN in most of the cases, even if they would better adapt to areas with low point density. Conversely, the context extension allows to compensate for the small block dimension and renders these modes competitive.
% as the partitioning algorithm aims to reduce as much as possible empty blocks while maximizing the performance of VoxelDNN, with context extension, the encoder will not prefer to encode block 64 as a single block but partition it to lower resolution to remove the sparsity while still having the context from block 64.
As a result, context extension significantly affects the optimal partitioning and enables VoxelDNN to adapt better to local sparsity while still providing enough contextual information to predict accurate probabilities.

\subsubsection{Using multiple models for the context} For the multi-resolution encoder, instead of using a separate model for each block size, VoxelDNN can use only a single neural network to predict the distribution. Specifically, we place small blocks (8, 16, 32) into a block of size 64 and then use the network for block 64 to predict and extract the corresponding distributions. This method of computing the occupancy distribution is different from Context Extension in that the surrounding voxels are always set to $0$. In Table \ref{table:singlevsmulti}, we compare the performance of using a single model with \textbf{Baseline}, which is a multi-models encoder. In this experiment, both encoders have 4 maximum partitioning levels and use the same model 64. On average, by having a separate model for each block size, a multi-model encoder obtains about $1\%$ additional gain over G-PCC compared to the single model encoder. This amount of gain indicates that the bigger VoxelDNN model can predict the conditional distribution on smaller blocks as efficiently as using a separate model for each block size. However, model 64 is trained on blocks of size 64 only, and learns features at that scale. In general, a model trained on small blocks could better capture the context from small input blocks and thus provides a higher gain in some circumstances.

\begin{table}[t]
\caption{Single model and multi-models comparison.}
\resizebox{0.97\linewidth}{!}{ \begin{tabular}{|P{0.1cm}|l|P{0.9cm}||R{0.9cm}|R{1.2cm}|R{0.9cm}|R{1.2cm}|}
\cline{3-7}
% \multicolumn{2}{|c||}{\begin{bf} Test PC \end{bf}}
\multicolumn{2}{c|}{}
& \begin{bf} G-PCC \end{bf}
& \multicolumn{2}{c|}{\begin{bf}Single model\end{bf}}
& \multicolumn{2}{c|}{\begin{bf}Multi-models\end{bf}}
\\
\cline{2-7}
\multicolumn{1}{c|}{}&Point Cloud & bpov&bpov&Gain over G-PCC &bpov&Gain over G-PCC\\
\hline

\multirow{3}{*}{\rotatebox[origin=c]{90}{MVUB}}&Phil& 1.1617 &08312 &-28.45\%& 0.8205 &-29.37\%  \\ 
\cline{2-7}
&Ricardo&1.0672 &0.7541 &-29.34\%& 0.7440&-30.28\% \\
\cline{2-7}
&\textbf{Average}  &\textbf{1.1145}  &\textbf{0.7927} &\textbf{-28.89\%} & \textbf{0.7823}&\textbf{-29.83\%}  \\
\cline{2-7}
\hline

\multirow{5}{*}{\rotatebox[origin=c]{90}{8i}}&Redandblack &1.0899  &0.7320 &-32.84\%& 0.7190&-34.3\% \\
\cline{2-7}
&Loot &0.9524&0.6403 &-32.77\% &0.6271 & -34.16\%\\
\cline{2-7}
&Thaidancer&0.9985 &0.7305 &-26.84\%&0.7297 &-26.92\% \\
\cline{2-7}
&Boxer&0.9479  &0.6008 &-36.62\%& 0.5900&-37.76\% \\
\cline{2-7}
&\textbf{Average} &\textbf{0.9972} &\textbf{0.6759}& \textbf{-32.27\%}&\textbf{0.6665} &\textbf{-33.22\%} \\
\hline

\multirow{4}{*}{\rotatebox[origin=c]{90}{CAT1}}&Frog &1.9085   &1.8433 &-3.42\%& 1.8214&-4.56\%  \\
\cline{2-7}
&Arco Valentino &4.8119 &5.2173 &+8.42\%& 5.2050&+8.17\% \\
\cline{2-7}
&Shiva&3.6721 & 3.6595 &-0.34\%&3.6403&-0.87\% \\
\cline{2-7}
&\textbf{Average} &\textbf{3.4642} &3.5734 &+1.56\%&\textbf{3.5556} &\textbf{+0.91\%}\\
\hline

\multirow{3}{*}{\rotatebox[origin=c]{90}{USP}}&BumbaMeuBoi&5.4522   &5.7501 &+5.46\%& 5.7305&+5.10\%  \\
\cline{2-7}
&RomanOiLight& 1.8604  &1.7094 &-8.12\%& 1.7030&-8.46\% \\
\cline{2-7}
&\textbf{Average}  &\textbf{3.6563}  &3.7298 &-1.33\%&\textbf{3.7168}&\textbf{-1.68\%}  \\
\cline{2-7}

\hline
\hline
\end{tabular}}
\label{table:singlevsmulti}

\end{table}

\subsubsection{Visualization of the bitrate allocation on coded PCs} The bpov heatmaps of 4 point cloud are shown in Figure \ref{fig:bitallocation}. The blocks in the figures reflect the optimal partitioning obtained by the algorithm. 
%All voxels within the same block have the same bpov value as we compute bpov for each output block of the partition.
First, we visually confirm what found in Figure~\ref{fig:bpovperformace}, i.e., VoxelDNN performs better, i.e., achieves a small bitrate, in the smooth and dense areas of the point cloud. Conversely, in the noisy areas  (\textit{Phil}'s hand, \textit{Loot}'s hand), sudden holes (\textit{Arco Valentino}) or very sparse regions (edges in \textit{Arco Valentino}, the bottom part of \textit{BumbaMeuBoi}), which are indicated in red, the performance is worse. 
% Overall, our approach is efficient on  dense and smooth content (e.g., Phil and Loot). BumbaMeuBoi is smooth and does not have much noise but it is a sparse point cloud, especially bottom part therefore, the performance is worser compare to Phil and Loot. On the contrary, noises, sparsity and rough surfaces exist almost on Arco Valentino and thus the bpov is worst. 
We can argue that the density of a point cloud, together with the smoothness and noise characteristics of the content, are among the main factors that influence the performance of VoxelDNN. On the other hand, we can argue that noisy and very sparse areas are intrinsically difficult to code in general, and indeed also the MPEG G-PCC codec requires a large number of bits to encode point clouds such as \textit{BumbaMeuBoi} and \textit{Arco Valentino}.

\subsection{Computational complexity analysis} A well-known drawback of auto-regressive generative models such as PixelCNN and VoxelDNN is the sequential generation of the symbol probabilities. This requires to run the network for each voxel, which has a complexity that increases linearly with the number of voxels. Therefore, VoxelDNN has a computational complexity which is 3 orders of magnitures bigger than G-PCC.

\begin{table}[t]
\caption{Average runtime (in seconds) of different encoders comparing with G-PCC.}
\centering
\resizebox{0.97\linewidth}{!}{ \begin{tabular}{M{0.7cm}M{1.2cm}M{1.5cm}M{1.8cm}}
\hline
\begin{bf}  \end{bf}
&\begin{bf}G-PCC\end{bf}
%&\begin{bf}1 level\end{bf}
%& \begin{bf}2 levels\end{bf} 
%&\begin{bf}3 levels\end{bf}
& \begin{bf}Baseline\end{bf} 
& \begin{bf}Baseline + CE\end{bf} \\
\hline
(Enc) &2.91&  3282&9271\\
(Dec) &2.85 &6783 &5765\\
\hline
\end{tabular}}
\label{table:complexity}
\end{table}

Table \ref{table:complexity} reports the encoding and decoding time of our \textbf{Baseline} and \textbf{Baseline + CE}. 
Tests are benchmarked on an Intel(R) Xeon(R) Silver 4110 CPU @ 2.10GHz machine with an Nvidia GeForce GTX 2080 GPU and 16 GB of RAM, running Ubuntu 16.04. Our encoding time is highly dependent on the number of blocks and the number of voxels within each block. Besides, the number of modes in the partitioning algorithm and context extension also influence the complexity. 
% Since our implementations are mainly for evaluation purposes, we only serve complexity analysis as the reference for the complexity of our method.
The \textbf{Baseline + CE} encoder tries all the extending modes and selects the best one, thus its average encoding time is higher than \textbf{Baseline} -- an increase of about 182\%. The reason why the encoding time for the \textbf{Baseline} codec is lower than the decoding time is purely implementative: at the encoder it is possible to predict the whole block probabilities in a single batch on a GPU, while in a realistic scenario, at the decoder side the voxels need to be individually decoded. 
When context extension is enabled, point clouds are partitioned into even smaller blocks, corresponding to a smaller complexity at the decoder, as a smaller number of voxels need to be decoded. On the other hand, the total parameters of each VoxelDNN model corresponds only to about 3.5 MB which is a small-size network in practice. Notice that the bottleneck in our system comes from the adoption of an auto-regressive model, which has the advantage of providing, in principle, an exact likelihood estimation of the data, though at a high computational cost. We are currently investigating the use of alternative generative approaches that avoid sequential probability estimation.

\section{Conclusions and future work}
\label{conclusion}

\par This paper presents a lossless compression method for point cloud geometry. We extend a well-known auto-regressive generative model initially proposed for 2D images to the 3D voxel space, and we incorporate  3D data augmentation to efficiently exploit the redundancies between points. This approach enables to build accurate probability models for the arithmetic coder. As a result, when using an adaptive partitioning scheme and context extension, our solution outperforms MPEG G-PCC over a diverse set of point clouds. 

Our analyses on the performance of the proposed method indicate at least two major avenues for improvement. On one hand, handling low-density point clouds would require to rethink the network architecture to handle sparse input data. On the other hand, a major drawback of VoxelDNN is the high computational cost of sequential probability generation, which we plan to replace in the future by a more efficient generative model.

\ifCLASSOPTIONcaptionsoff
  \newpage
\fi

\bibliographystyle{./IEEEtran}
\bibliography{./IEEEabrv,./refs}

\end{document}